\documentclass[PAPER,cernpreprint, atlasdraft=false, UKenglish, texmf, orcidlogo]{atlasdoc}
 
\usepackage[subfigure]{atlaspackage}
\usepackage{atlasbiblatex}
 
\usepackage{atlasphysics}

\graphicspath{{logos/}{figures/}}
 
\usepackage{ANA-HDBS-2018-52-PAPER-defs}

\AtlasTitle{Searches for exclusive Higgs boson decays into $D^*\gamma$ and $Z$ boson decays into $D^0\gamma$ and $K^0_s\gamma$ in $pp$ collisions at $\sqrt{s}=13\,\textrm{TeV}$ with the ATLAS detector}

\AtlasAbstract{
Searches for exclusive decays of the Higgs boson into $D^*\gamma$ and of the $Z$ boson into $D^0\gamma$ and $K^0_s\gamma$ can probe flavour-violating
Higgs boson and $Z$ boson couplings to light quarks.
Searches for these decays are performed with a $pp$ collision data sample corresponding to an integrated luminosity of $136.3$~fb$^{-1}$ collected at $\sqrt{s}=13\,\textrm{TeV}$ between 2016--2018 with the ATLAS detector at the CERN Large Hadron Collider. In the $D^*\gamma$ and $D^0\gamma$ channels, the observed (expected) 95$\%$ confidence-level
upper limits on the respective branching fractions are
${\cal B}(H\rightarrow D^*\gamma)< 1.0 (1.2)\times 10^{-3}$,
${\cal B}(Z\rightarrow D^0\gamma)< 4.0 (3.4)\times 10^{-6}$, while the corresponding results in the $K^0_s\gamma$ channel are ${\cal B}(Z\rightarrow K^0_s\gamma)< 3.1 (3.0)\times 10^{-6}$.
}

\AtlasRefCode{HDBS-2018-52}
 
\PreprintIdNumber{CERN-EP-2024-037}

\AtlasJournalRef{\PLB 855 (2024) 138762}
\AtlasDOI{https://doi.org/10.1016/j.physletb.2024.138762}

\hypersetup{pdftitle={ATLAS document},pdfauthor={The ATLAS Collaboration}}
 
\begin{document}
 
\maketitle

\section{Introduction}
\label{sec:intro}
 
After the observation of the Higgs boson ($H$) with a mass of 125~\GeV by the ATLAS~\cite{PERF-2007-01} and CMS~\cite{CMS-TDR-08-001} Collaborations~\cite{HIGG-2012-27,CMS-HIG-12-028}, many studies were performed to measure its properties which, so far, are consistent with Standard Model (SM) expectations~\cite{HIGG-2021-23_werratum,CMS-HIG-22-001}.
These have confirmed its role in the spontaneous breaking of electroweak symmetry and the mass generation of the massive vector bosons~\cite{Englert:1964et,Higgs:1964ia}.
A complete observation of the Higgs boson Yukawa couplings to third-generation charged fermions was achieved by the ATLAS and CMS collaborations through the observation of
the decays $H\rightarrow\tautau$~\cite{HIGG-2019-09,CMS-HIG-19-010} and $H\to\bbbar$~\cite{HIGG-2018-51,CMS:2023vzh},
and the production of Higgs bosons with top-quark pairs~\cite{HIGG-2018-13,CMS-HIG-17-035}.
Evidence was also reported for the decay $H\to\mumu$~\cite{CMS-HIG-19-006,HIGG-2019-14}, and direct searches for $H\to\ccbar$~\cite{HIGG-2021-12,CMS-HIG-21-008} and $H\to\ee$ decays~\cite{HIGG-2018-58,CMS-HIG-13-007} were performed, but  there is no further experimental evidence for the Higgs boson couplings to the first and second generations of fermions.
Instead, the light ($u$, $d$, $s$) quark couplings to the Higgs boson are loosely constrained by existing data on the total Higgs boson's width and combined measurements of Higgs boson production and decays~\cite{HIGG-2021-23_werratum,CMS-HIG-22-001}.
 
The ATLAS and CMS Collaborations have also investigated potential beyond-the-SM (BSM) couplings of the Higgs boson,
including searches for the lepton-flavour-violating decays $H\to e\mu$, $H\to e\tau$ and $H\to \mu\tau$~\cite{HIGG-2018-58,HIGG-2019-11,HIGG-2017-08,CMS-HIG-20-009}
and for flavour-changing neutral currents via the $t$-quark decays  $t\rightarrow cH$ and $t\rightarrow uH$~\cite{TOPQ-2019-04_pub,TOPQ-2017-15,CMS-TOP-19-002,CMS-TOP-20-007-PRL}.
An overview of BSM theories which allow flavour-violating couplings of the Higgs boson to quarks is presented in Ref.~\cite{deFlorian:2016spz}.
These include BSM scenarios such as the minimal flavour violation framework~\cite{DAmbrosio:2002ex} and the Giudice--Lebedev Higgs boson dependent Yukawa couplings model~\cite{Giudice:2008uua}.
 
\begin{figure}[!htbp]
\centering
\includegraphics[width=0.4\textwidth]{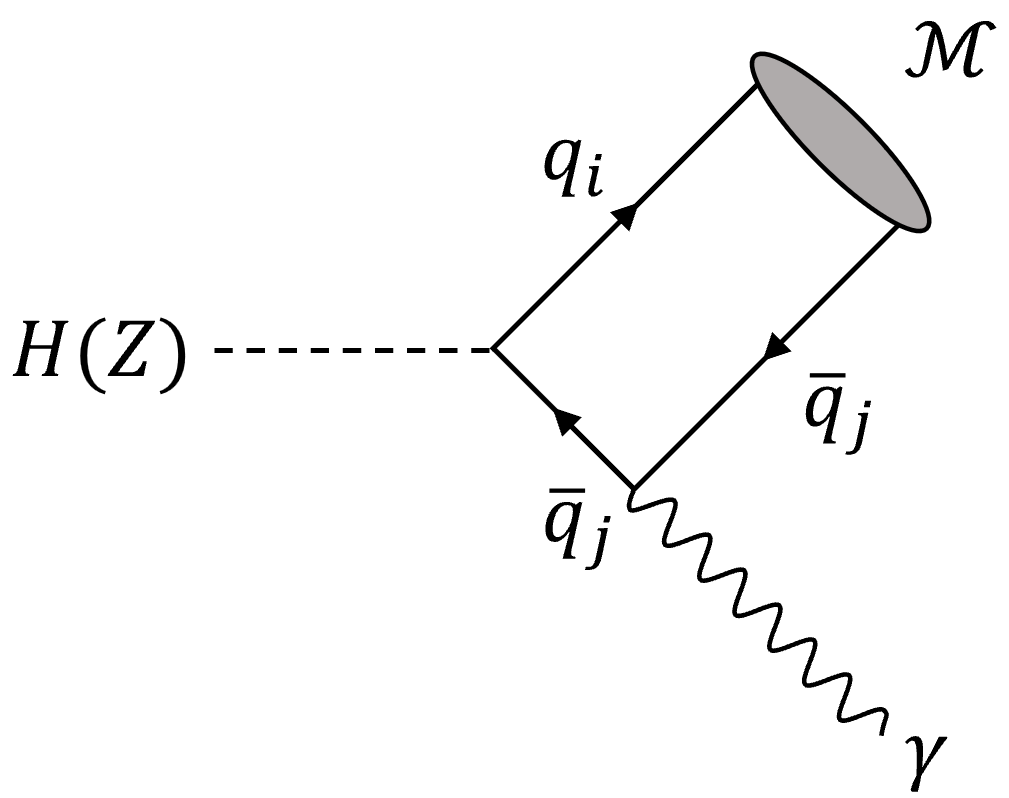}\label{fig:dispVtx:flavourChangingAmp}
\caption{An illustrative Feynman diagram depicting the flavour-violating $H\rightarrow \mathcal{M}\gamma$ and $Z\rightarrow \mathcal{M}\gamma$ processes considered in this search, where $\mathcal{M}$ is a flavoured meson.
For Higgs boson decays, $\mathcal{M} = D^*$; for $Z$ boson decays, $\mathcal{M}=D^0,K^0_s$.
The indices $i$ and $j$ refer to the flavour of the quark, and $i \neq j$.
\label{fig:dispVtx:feynman}}
\end{figure}

Rare exclusive decays of the Higgs boson into a meson and a photon  probe both the potential flavour-violating Higgs boson couplings and the Yukawa couplings of the SM~\cite{Kagan:2014ila,HBranching,VatoHBranching,Koenig:2015pha,Bodwin:2014bpa,Bodwin:2016edd,Bodwin:2017wdu,Zhou:2016sot,Brambilla:2019fmu}.
Analogous decays of the $Z$ boson into a meson and a photon offer unique tests of the factorisation approach in quantum chromodynamics, QCD~\cite{Huang:2014cxa,Grossmann:2015lea,Bodwin:2017pzj}, and in particular,
probes of potential flavour-changing neutral current interactions of the $Z$ boson via such decays are discussed in Ref.~\cite{Grossmann:2015lea}. Searches for these exclusive decays were performed by the ATLAS and CMS Collaborations~\cite{HIGG-2014-03,HIGG-2016-23,HDBS-2018-53-new,HIGG-2016-05,HIGG-2016-13_werratum,HDBS-2019-33-new,CMS-HIG-14-003,CMS-SMP-17-012},
and the ATLAS constraints on them, with corresponding SM predictions, are summarised in Ref.~\cite{ATL-PHYS-PUB-2023-004}.
Figure~\ref{fig:dispVtx:feynman} shows an illustrative Feynman diagram of the $H(Z)\rightarrow\mathcal{M}\gamma$ process (where $\mathcal{M}$ denotes a meson) which proceeds via a flavour-violating coupling
that can be probed with searches for exclusive Higgs bosons and $Z$ boson decays into a meson and a photon,
as done previously in the ATLAS search for $H\to K^*\gamma$~\cite{HDBS-2019-33-new}.
 
Motivated by the potential for flavour-violating interactions in the couplings of the Higgs and $Z$ bosons to quarks,  searches are described for the decays $H\rightarrow D^*\gamma$, $Z\rightarrow D^0\gamma$ (and the corresponding charge-conjugate decays $H\rightarrow \overline{D}{}^{*}\gamma$ and $Z\rightarrow \overline{D}{}^{0}\gamma$) and $Z\rightarrow K^0_s\gamma$, which use $136.3~\ifb$ of ATLAS $pp$ collision data collected at $\sqrt{s} = 13~\TeV$. For the $D^*\gamma$, $D^0\gamma$ and $K^0_s\gamma$ final states, only the three previously mentioned specific decay channels are allowed by angular momentum conservation.
Hereafter, $D^*$ includes the $D^*$ and its anti-particle $\overline{D}{}^{*}$, and $D^0$ includes the $D^0$ and its anti-particle $\overline{D}{}^{0}$.
 
In the SM, the $D^{*}\gamma$ decay arises only from loop contributions. The SM branching fraction for Higgs boson decays into $\bar{c}u$ or $\bar{u}c$ has been calculated to be $\mathcal{B}(H\to \bar{c}u/\bar{u}c)=5\times10^{-20}$~\cite{Aranda:2020tqw}, while the branching fraction for $H\to D^*\gamma$ is expected to be much smaller, estimated at $7\times10^{-27}$~\cite{dEnterria:2023wjq}. The corresponding SM branching fractions for the $Z$ boson decays are similarly small~\cite{Grossmann:2015lea,dEnterria:2023wjq}, estimated at $1.4\times10^{-25}$ and $3.3\times10^{-20}$ for the $Z\rightarrow D^0\gamma$ and $Z\rightarrow K^0_s\gamma$ decays, respectively~\cite{dEnterria:2023wjq}.
The decays $H\rightarrow D^*\gamma$ and $Z\rightarrow D^0\gamma$ probe for flavour-violating couplings of the Higgs and $Z$ bosons to $u$- and $c$-quarks, and the decay $Z\rightarrow K^0_s\gamma$ probes for flavour-violating couplings of the $Z$ boson to $d$- and $s$-quarks.
The LHCb experiment has searched for the decay $Z\rightarrow D^0\gamma$ using $2.0~\ifb$ of data collected at $\sqrt{s}=13$~\TeV, yielding a $95\%$ CL upper limit of $2.1\times 10^{-3}$~\cite{LHCb:2022kta}, but there are no further experimental constraints on these decays.
Given that the expected SM branching fractions for these decays are vanishingly small, an observation could imply physics beyond the SM, such as the existence of these flavour-violating couplings.
 
The signatures of the exclusive decays into a meson and a photon include a high-energy photon and a meson appearing approximately back-to-back in the detector following the decay of the $H$ or $Z$ boson, where there is a resonance in di-track mass to reconstruct the meson and in three-body mass to reconstruct the initial boson~\cite{HIGG-2014-03,HIGG-2016-23,HDBS-2018-53-new,HIGG-2016-05,HIGG-2016-13_werratum,HDBS-2019-33-new,CMS-HIG-14-003,CMS-SMP-17-012}.
The decays considered in this paper have an additional feature of a displaced decay vertex, either through the decay of the $D^0$ in $H\rightarrow D^*\gamma$ and $Z\rightarrow D^0\gamma$, or the decay of the $K^0_s$ in $Z\rightarrow K^0_s\gamma$.
This provides a particularly distinct signature as requirements on the vertex displacement are used to reject multijet events, the dominant contribution to the background.
The backgrounds in these searches are considered inclusively, and originate primarily from multijet and $\gamma$+jet events involving the production of the meson or a non-resonant di-track system near its mass.
The two primary decay channels of the $D^*$ are considered, into $D^0\pi^0$ and $D^0\gamma$, where the $\pi^0$ and photon are soft and are not explicitly reconstructed.
The searches for $H\to D^*\gamma$ and $Z\to D^0\gamma$ are grouped into a single event selection which targets the decay $D^0\to K^-\pi^+$, while the search for $Z\to K^0_s\gamma$ has a separate selection which targets the decay $K^0_s\to\pipi$.


\section{ATLAS detector}
\label{sec:detector}
 
The ATLAS detector~\cite{PERF-2007-01} is a multipurpose particle
physics detector with an approximately forward--backward symmetric cylindrical
geometry and near $4\pi$ coverage in solid angle.\footnote{ATLAS uses
a right-handed coordinate system with its origin at the nominal
interaction point (IP) in the centre of the detector and the $z$-axis
along the beam pipe. The $x$-axis points from the IP to the centre of
the LHC ring, and the $y$-axis points upward. Cylindrical coordinates
$(r, \phi)$ are used in the transverse plane, $\phi$ being the
azimuthal angle around the $z$-axis. The pseudorapidity is defined in
terms of the polar angle $\theta$ as $\eta = -
\ln{\tan(\theta/2)}$.} It consists of an inner tracking
detector surrounded by a thin superconducting solenoid,
electromagnetic and hadronic calorimeters, and a muon spectrometer.
The inner tracking detector (ID) covers the pseudorapidity range of
$|\eta| < 2.5$, and is surrounded by a thin superconducting solenoid
providing a 2~T magnetic field. At small radii, a high-granularity
silicon pixel detector covers the vertex region and typically provides
three measurements per track. A new innermost pixel-detector layer,
the insertable B-layer, was added before 13~\TeV\ data-taking began in
2015 and provides an additional measurement at a radius of about 33~mm
around a new and thinner beam pipe~\cite{ATLAS-TDR-19,PIX-2018-001}. The pixel
detectors are followed by a silicon microstrip tracker, which
typically provides four space-point measurements per track. The
silicon detectors are complemented by a gas-filled straw-tube
transition radiation tracker, which enables radially extended track
reconstruction up to $|\eta| = 2.0$, with typically 35 measurements
per track.
The calorimeter system covers the pseudorapidity range of $|\eta| <
4.9$. A high-granularity lead/liquid-argon (LAr) sampling
electromagnetic calorimeter covers the region $|\eta| < 3.2$, with an
additional thin LAr presampler covering $|\eta| < 1.8$ to correct for
energy losses upstream. The electromagnetic calorimeter is divided
into a barrel section covering $|\eta| < 1.475$ and two endcap
sections covering $1.375 < |\eta| < 3.2$. For $|\eta| < 2.5$ it is
divided into three layers in depth, which are finely segmented in
$\eta$ and $\phi$. A steel/scintillator-tile calorimeter provides
hadronic calorimetry in the range of $|\eta| < 1.7$, while in the endcap
region, $1.5 < |\eta| < 3.2$, a copper/LAr calorimeter is used. The
solid-angle coverage is completed with forward copper/LAr and
tungsten/LAr calorimeter modules in $3.1 < |\eta| < 4.9$, optimised
for electromagnetic and hadronic measurements, respectively.
The muon spectrometer surrounds the calorimeters and comprises
separate trigger and high-precision tracking chambers measuring the
deflection of muons in a magnetic field provided by three air-core
superconducting toroids.
 
A two-level trigger and data acquisition system is used to provide
online selection and record events for offline
analysis~\cite{TRIG-2016-01}. The level-1 trigger is implemented in
hardware and uses a subset of detector information to reduce the event
rate to 100~kHz or less from the maximum LHC collision rate of 40~MHz.
It is followed by a software-based high-level trigger which
filters events using the full detector information and records events
for detailed offline analysis at an average rate of 1~kHz.
An extensive software suite~\cite{ATL-SOFT-PUB-2021-001} is used in
data simulation, in the reconstruction and analysis of real and
simulated data, in detector operations, and in the trigger and data
acquisition systems of the experiment.


\section{Data and simulation}
\label{sec:dataMC}
 
The searches are performed with a sample of $pp$ collision data collected at a centre-of-mass energy $\sqrt{s} = 13$~\TeV from 2016--2018.
The total integrated luminosity available is 136.3~fb$^{-1}$ for each final state, following the requirement that events must be collected under stable LHC beam conditions with all relevant detector components in good operating condition~\cite{DAPR-2018-01}.
 
The data samples were recorded by a combination of dedicated triggers that were available from May 2016 and remained operational until the end of data taking in 2018.
The search for $H\to D^*\gamma$ and $Z\to D^0\gamma$ uses triggers which target $D^0\to K^-\pi^+$ decays whilst the search for $Z\to K^0_s\gamma$ uses triggers which target $K^0_s\to\pipi$ decays.
These decay channels were selected as they each involve exactly two charged hadrons, such that the meson candidates can be fully reconstructed, thus providing  favourable information for triggering.
Each trigger requires a photon at the level-1 trigger with $\pt > 24$~\GeV.
At the high-level trigger, an isolated photon with a transverse momentum $\pt^{\gamma} > 35$~\GeV~\cite{TRIG-2018-05} is required for the 2016 data taking period, and with a reduced threshold of $\pt^{\gamma} > 25$~\GeV throughout 2017--2018.
For the ID tracks required at the high-level trigger, modified versions of the $\tau$-lepton trigger algorithms~\cite{ATLAS-CONF-2017-061} are used.
Each trigger requires a pair of tracks that is matched to a topological cluster of calorimeter cells~\cite{PERF-2014-07}
with a transverse energy greater than 25~\GeV.
Within the pair, one track is required to have $\pt$ greater than 15~\GeV.
Different requirements on the invariant mass of the pair of tracks are applied, depending on the targeted meson decay.
For $D^0\to K^-\pi^+$ an invariant mass of the pair of tracks in the range 1800--1930~\MeV\ is required under the $K^{\pm}\pi^{\mp}$ hypothesis,
whilst for $K^0_s\to \pipi$ an invariant mass in the range 460--538~\MeV is required under the charged-pion hypothesis. 
 
The generation and normalisation of simulated signal samples follow the methods used in the search for $H\to K^*\gamma$ and $H(Z)\to\omega\gamma$~\cite{HDBS-2019-33-new} and are summarised here.
Higgs boson production through gluon ($ggH$) and vector-boson fusion (VBF) processes was modelled up to next-to-leading-order (NLO) in $\alphas$
using the \POWHEGBOX[v4] Monte Carlo (MC) event generator~\cite{powheg1,powheg2,powheg3,powheg4,powheg5}. \POWHEGBOX[v4] was interfaced
with the \Pythia[8.244] MC event generator~\cite{Sjostrand:2007gs} to model the parton shower, hadronisation and underlying event, with
parameter values set according to the AZNLO set of tuned parameters (``tune'')~\cite{STDM-2012-23} and using CTEQ6L1 parton distribution functions (PDFs)~\cite{Pumplin:2002vw}.
Additional contributions from the  production of a Higgs boson with a $W^{\pm}$ or $Z$ boson (denoted by $WH$ and $ZH$, respectively) were also modelled with \POWHEGBOX[v4], but interfaced to \Pythia[8.244]  with \NNPDF[2.3lo] PDFs~\cite{Ball:2012cx} and the A14 tune~\cite{ATL-PHYS-PUB-2014-021} for hadronisation and the underlying event. Higgs boson production  with top quarks ($t\bar{t}H$) was modelled at NLO using the  event generator \AMCatNLO~\cite{Alwall:2014hca} interfaced to \Pythia[8.244], again with the \NNPDF[2.3lo] PDFs and A14 tune. Events were generated assuming a Higgs boson mass of $m_H = 125$~\GeV, but were normalised to cross-sections associated with $m_H = 125.09$~\GeV. These were obtained from Ref.~\cite{LHCHCWG:2012vm} and are summarised below.
The $ggH$ production rate was normalised such that it reproduces the total cross-section predicted by a next-to-next-to-next-to-leading-order QCD calculation with NLO electroweak corrections applied~\cite{Actis:2008ug,Anastasiou:2008tj}.
The VBF production rate was normalised to an approximate next-to-next-to-leading-order (NNLO) QCD cross-section with NLO electroweak corrections applied~\cite{Ciccolini:2007jr,Ciccolini:2007ec,Bolzoni:2010xr}.
The $WH$ and $ZH$ production rates were normalised to cross-sections calculated at NNLO in QCD with NLO electroweak corrections~\cite{Brein:2003wg} including the NLO QCD
corrections for $gg\to ZH$. \POWHEGBOX[v4] was also used to
model inclusive $Z$ boson production with CT10 PDFs~\cite{Lai:2010vv}. \Pythia[8.244] with CTEQ6L1 PDFs~\cite{Pumplin:2002vw} and
the AZNLO tune was used to simulate the parton showering and hadronisation. The prediction is
normalised to the total cross-section obtained from the measurement in Ref.~\cite{STDM-2015-03}.
 
The Higgs and $Z$ boson decays were simulated as a cascade of two-body decays.
The branching fractions for $D^*\to D^0\pi^0$ and $D^*\to D^0\gamma$ are $(64.7\pm0.9)\%$ and $(35.3\pm0.9)\%$ respectively~\cite{PDG}.
The branching fraction for the decay $D^0\to K^-\pi^+$ is $(3.947\pm0.030)\%$ and the branching fraction for the decay $K^0_s\to\pipi$ is $(69.20\pm 0.05)\%$.
The natural lifetime of the $D^0$ meson is $(4.103\pm 0.010)\times 10^{-13}$~s and for the $K^0_s$ meson is $(8.954\pm0.004)\times 10^{-11}$~s.
The simulated events were passed through a detailed \GEANT simulation of the ATLAS detector~\cite{GEANT4,2010wqa} and processed with the same software used to reconstruct the data.
The generation of the simulated event samples includes the effect of multiple \(pp\) interactions per bunch crossing, and the effect on the detector response due to interactions from bunch crossings before or after the one containing the hard interaction.


\section{Physics objects and event selection}
\label{sec:selection}
 
In addition to the trigger and data-quality requirements, several selection criteria are defined to retain events for further analysis.
The searches are split into two distinct event selections that share several common requirements.
The first selection, henceforth called the $D^0\gamma$ selection, covers the search for both the $H\to D^*\gamma$ and $Z\to D^0\gamma$ decays and targets the reconstruction of the $D^0$ meson.
The additional soft daughter pion or photon from the decay of the $D^*$ is not explicitly reconstructed;
this does not deteriorate the sensitivity to the $H\to D^*\gamma$ signal. The second selection, the $K^0_s\gamma$ selection, covers the search for the $Z\to K^0_s\gamma$ signal and targets the reconstruction of the $K^0_s$ meson.
 
Events with a $pp$ interaction vertex reconstructed from at least two ID tracks with $\pt > 500~\MeV$ are considered in the analysis. Within an event, the primary vertex is defined as the reconstructed vertex with the largest $\sum\pt^{2}$ of associated ID tracks. Common to each event selection are the photon and charged-hadron requirements.
Photons are reconstructed from clusters of energy in the electromagnetic calorimeter,
and clusters which match ID tracks consistent with the hypothesis of a photon conversion into $\ee$ are classified as converted photon candidates,
whilst clusters which have no matching ID tracks are classified as unconverted photon candidates~\cite{EGAM-2018-01}.
Photon candidates are required to satisfy \enquote{tight} photon identification criteria~\cite{EGAM-2018-01},
and have the kinematic and geometric requirements of $\pt^{\gamma} > 35$~\GeV and $|\eta^{\gamma}| < 2.37$, excluding the barrel-to-endcap calorimeter transition region $1.37 < |\eta^{\gamma}| < 1.52$.
Track- and calorimeter-isolation requirements are imposed to further suppress contamination from jets.
For the track isolation, the sum of the transverse momenta of all tracks within $\Delta R = \sqrt{(\Delta\phi)^2 + (\Delta\eta)^2} = 0.2$ of the photon direction, excluding those matched to  the reconstructed photon itself, is required to be less than 5$\%$ of $\pt^{\gamma}$.
To mitigate the effects of multiple $pp$ interactions in the same or neighbouring bunch crossings (pile-up events), only ID tracks consistent with originating from the primary vertex are considered~\cite{PERF-2015-01}.
For the calorimeter isolation, the sum of the transverse momenta of all calorimeter energy deposits within $\Delta R = 0.4$ of the photon direction, excluding those matched to the reconstructed photon, is required to be less than $2.45~\GeV + 0.022 \times \pt^{\gamma}~[\GeV]$.
The effects of the underlying event and pile-up events are also accounted for on an event-by-event basis using an average underlying-event energy density determined from data~\cite{EGAM-2018-01}.
 
Charged-hadron candidates are reconstructed from ID tracks and are required to satisfy \enquote{loose} track identification criteria~\cite{ATL-PHYS-PUB-2015-051},
and require $\pt > 5$~\GeV and $|\eta| < 2.5$.
Pairs of charged-hadron candidates are combined to form the meson candidate $\cal{M}$ for each selection, and must have opposite charges.
For the $D^0\gamma$ final state, $\cal{M}$ is a $D^0$ meson candidate, and for the $K^0_s\gamma$ final state, $\cal{M}$ is a $K^0_s$ meson candidate.
The charged-hadron candidate in a pair with the higher $\pt$, the leading track, must satisfy $\pt > 20$~\GeV.
Meson candidates must satisfy the isolation requirement that the sum of the $\pt$ of the reconstructed ID tracks originating from the primary vertex within $\Delta R = 0.2$ of the leading charged-hadron candidate, excluding the pair of tracks which define the meson candidate itself, must be less than 10$\%$ of the $\pt$ of the meson candidate.
 
Combinations of meson and photon candidates must also satisfy $\Delta\phi > \pi/2$, where $\Delta\phi$ is the difference between the  azimuthal angle of the two candidates.
If an event has multiple photon candidates, the candidate with highest $\pt$ is chosen; if the event has multiple meson candidates, the candidate with mass closest to the target meson is selected.
However, these situations arise in a small fraction of data events: fewer than 5\% of events which satisfy the trigger have more than one meson candidate, and fewer than 0.5\% of events have more than one photon candidate.
 
The differences between the two event selections arise from requirements on $\cal{M}$ mass and $\pt$, and  requirements on the meson candidate vertex. In the $K^0_s\gamma$ channel, both the tracks are assigned the charged pion mass to calculate the $\cal{M}$ candidate mass. In the $D^0\gamma$ final state, given the absence of particle identification information in the relevant momentum range, the assignment of the charged kaon or pion masses which gives a value of the $\cal{M}$ mass closest to the $D^0$ mass is chosen. The $D^0\gamma$ final state requires that the meson candidate satisfies $1800$~\MeV $< m_{\cal{M}} < 1930$~\MeV and $\pt^{\cal{M}} > 39$~\GeV;
the $K^0_s\gamma$ final state requires $460$~\MeV $< m_{\cal{M}} < 538$~\MeV and $\pt^{\cal{M}} > 38$~\GeV.
To reject contributions from events with prompt vertices, requirements on $L_{xy}/\sigma_{L_{xy}}$, the $L_{xy}$ significance of the meson candidate vertex, are imposed.
Here $L_{xy}$ is the signed projection of the vector leading from the primary vertex to the meson candidate vertex onto the direction of the $\cal{M}$ candidate's $\pt$, and its corresponding uncertainty is $\sigma_{L_{xy}}$.
The $D^0\gamma$ final state requires $L_{xy}/\sigma_{L_{xy}} > 3$, whilst the $K^0_s\gamma$ final state requires $L_{xy}/\sigma_{L_{xy}} > 5$.
To remove events from particle interactions in the detector material, the radius of the meson candidate vertex must satisfy $r < 15$~mm for the $D^0\gamma$ selection, inside the beampipe, and $r < 65$~mm for the $K^0_s\gamma$ selection, between the second and third layers of the pixel detector. While many $K^0_s$ decays are expected to occur beyond this region, the trigger algorithm described in Section~\ref{sec:dataMC} is  very inefficient for such events and offline reconstructed vertices with $r > 65$~mm are dominated by material interactions not matched to the trigger signature. These selection criteria define the nominal \enquote{Signal Region} (SR) for each final state, and events which satisfy these criteria are retained for further analysis.


\section{Signal modelling}
\label{sec:signal}
 
Figure~\ref{fig:pt_ph_tracks} shows the generator-level $\pt$ distributions of the two tracks and the photon reconstructed in each of the signal decays before and after implementing the nominal SR event selection for each final state.
The trigger efficiency after the offline selection is $66\%$ for $H\to D^*\gamma$, $69\%$ for $Z\to D^0\gamma$, and $39\%$ for $Z\to K^0_s\gamma$ and is primarily due to differences between online and offline tracking performance.
For the $H\to D^*\gamma$ signals, the total signal efficiency, including kinematic acceptance, trigger and reconstruction efficiencies, is approximately $9\%$.
The corresponding total efficiency for $Z\to D^0\gamma$ is $3\%$ and for $Z\to K^0_s\gamma$ is $0.2\%$.
Overall, the lower efficiency to reconstruct the $Z$ boson decays is due to the softer $\pt$ distributions of the decay products.
The particularly small total efficiency for the $Z\to K^0_s\gamma$ signal is related to the large lifetime of the $K^0_s$ meson, as most decays occur beyond the inner layers of the ID, where the track reconstruction efficiency is significantly reduced.
 
\begin{figure}[!htbp]
\centering
\subfigure[]{\includegraphics[width=0.32\textwidth]{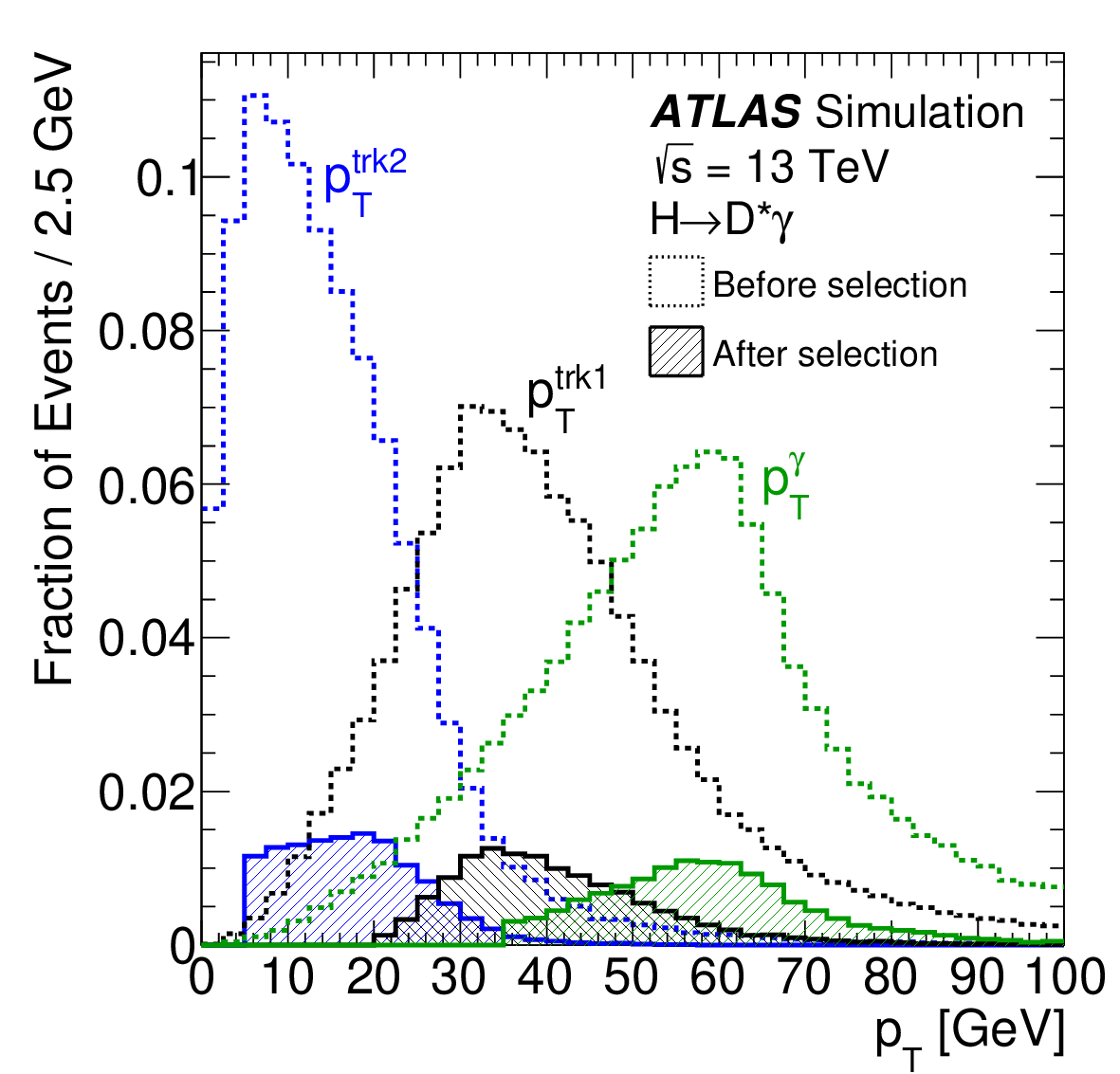}\label{fig:dzero:pt_ph_tracks_Hdstar}}
\subfigure[]{\includegraphics[width=0.32\textwidth]{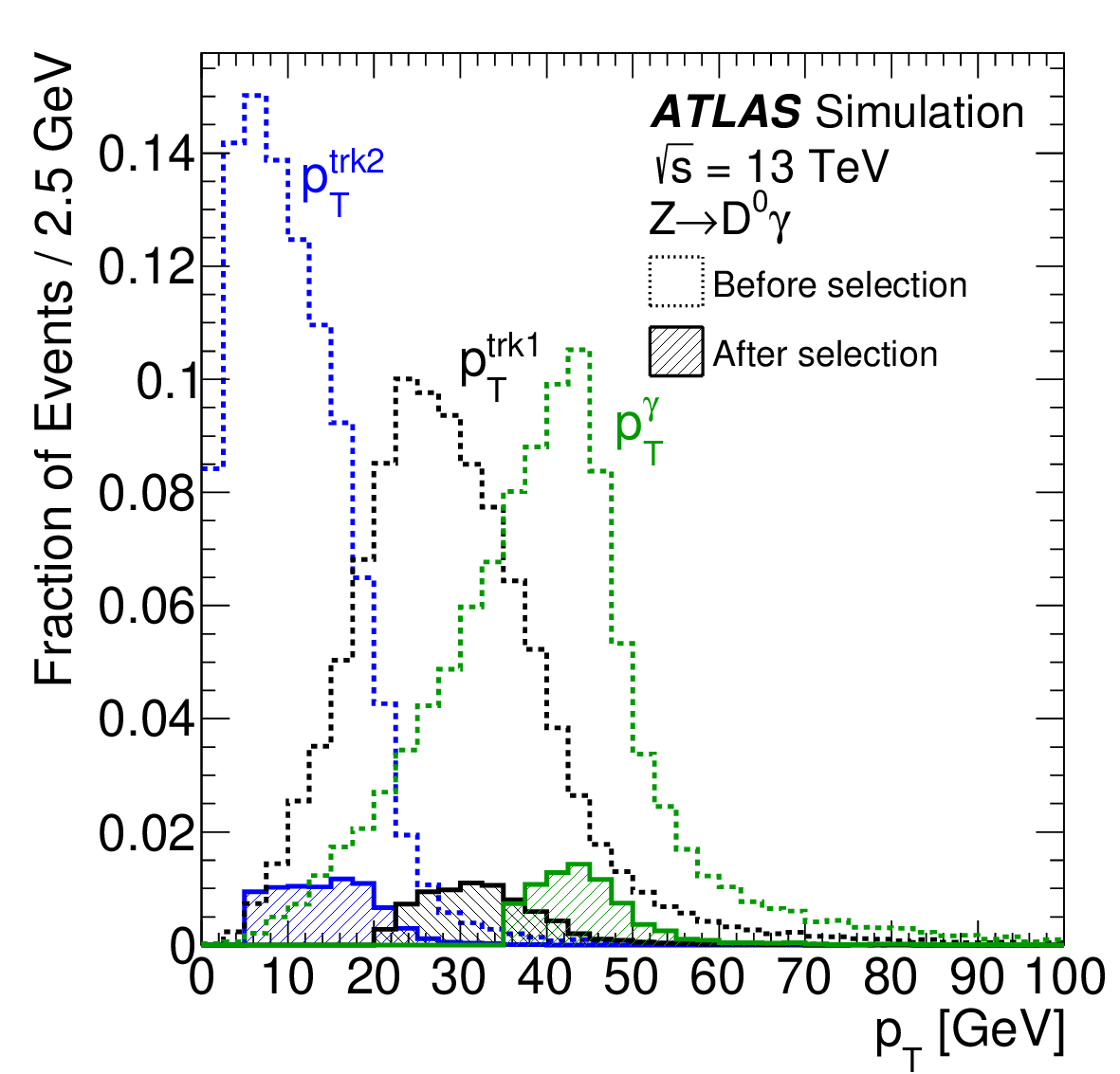}\label{fig:dzero:pt_ph_tracks_Zdzero}}
\subfigure[]{\includegraphics[width=0.32\textwidth]{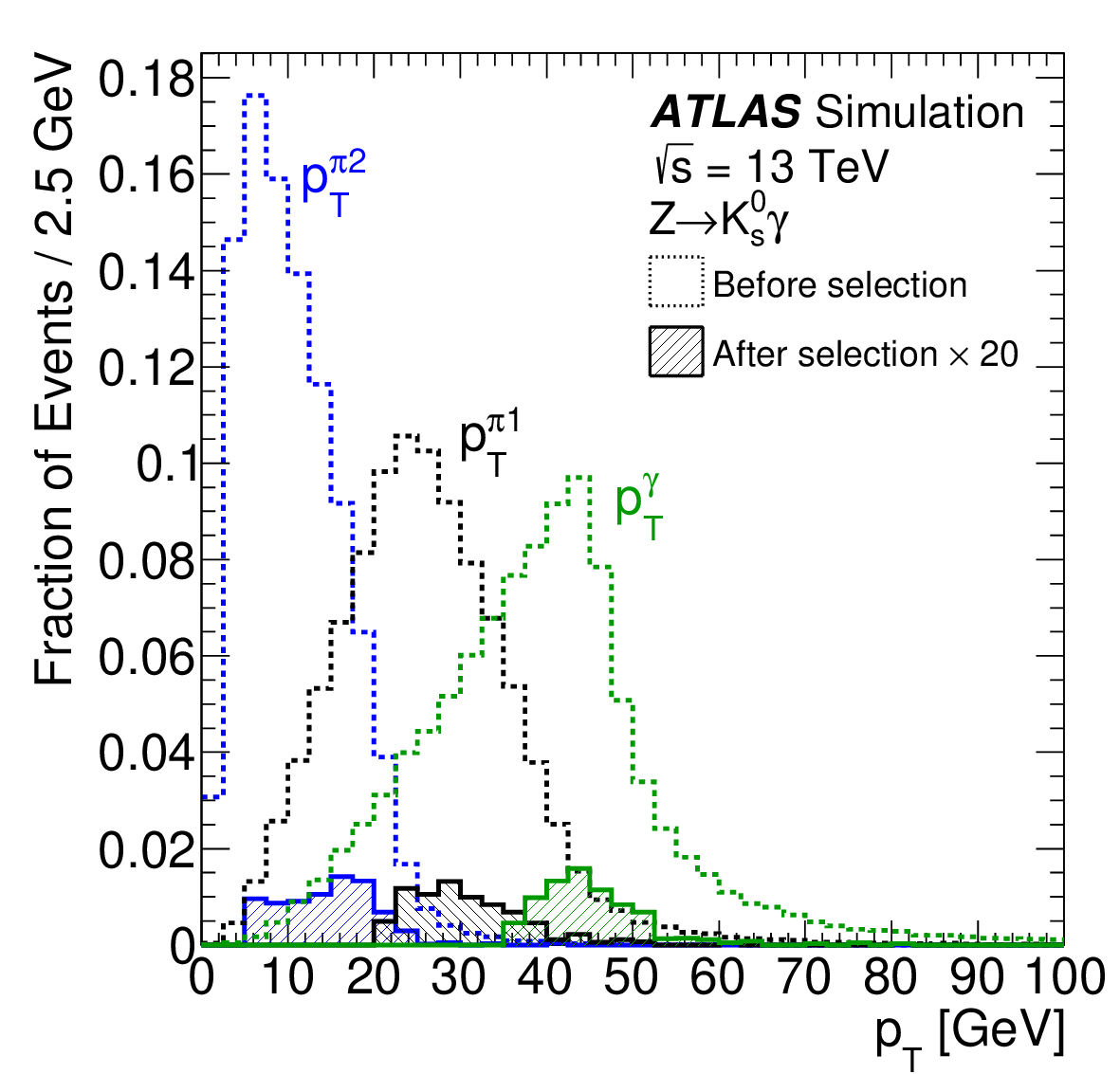}\label{fig:kshort:pt_ph_tracks_Zkshort}}\\
\caption{Generator-level transverse momentum ($\pt$) distributions of the photon and of the tracks, ordered in $\pt$,
for \subref{fig:dzero:pt_ph_tracks_Hdstar} $H\to D^{*}\gamma$, \subref{fig:dzero:pt_ph_tracks_Zdzero} $Z\to D^{0}\gamma$, and \subref{fig:kshort:pt_ph_tracks_Zkshort} $Z\to K^0_s\gamma$ simulated signal events.
The hatched histograms denote the full event selection while the dashed histograms show the events at generator level that fall in the
geometric acceptance (both the tracks are required to have $|\eta|<2.5$ while the photon is required to have $|\etagamma|<2.37$, excluding
the region $1.37<|\etagamma|<1.52$). The dashed histograms are normalised to unity, and the relative difference between
the two sets of distributions corresponds to the effects of reconstruction, trigger, and event selection efficiencies.
The leading track is denoted by $\pt^{\textrm{trk} 1}$ and the subleading candidate by $\pt^{\textrm{trk} 2}$ for $H\to D^{*}\gamma$ and $Z\to D^{0}\gamma$.
For $Z\to K^0_s\gamma$ these are labelled $\pt^{\pi 1}$ and $\pt^{\pi 2}$.\label{fig:pt_ph_tracks}}
\end{figure}
 
The Higgs boson mass distribution for $H\to D^*\gamma$ is modelled with a sum of two Gaussian probability density functions (pdf) with a common mean value.
The mass resolution achieved is $2.2\%$, and the mean is shifted to approximately $121$~\GeV without the reconstruction of the additional daughter $\pi^0$ or photon from the decay of the $D^*$.
For the $Z$ boson signals the $Z$ boson mass distribution is modelled with a Voigtian pdf, a convolution of relativistic Breit--Wigner and Gaussian pdfs, corrected with a mass-dependent efficiency factor which accounts for the turn on in signal efficiency versus $Z$ boson mass due to the kinematic requirements of the event selection.
The mass resolutions achieved are 1.9\% for $Z\to D^0\gamma$ and 2.3\% for $Z\to K^0_s\gamma$, and for each signal the mean of the Voigtian is approximately $91$~\GeV.


\section{Background modelling and validation}
\label{sec:background}
 
For each final state, the main sources of background are events involving inclusive multijet or $\gamma$\;\!+\,jet processes, where a meson candidate is reconstructed from ID tracks originating from a jet, and these
show a non-resonant kinematic structure in the three-body mass distribution.
The background processes are modelled with a non-parametric data-driven approach using very finely binned templates to describe the Higgs and $Z$ boson mass distributions~\cite{Chisholm:2021pdn}.
The procedure captures the correlations between the kinematic and isolation variables, thus both the multijet and $\gamma$\;\!+\,jet processes can be modelled inclusively.
This technique was also employed in previous ATLAS searches for exclusive decays of the Higgs and $Z$ bosons into a meson and a photon~\cite{HIGG-2014-03,HIGG-2016-23,HDBS-2018-53-new,HIGG-2016-05,HIGG-2016-13_werratum,HDBS-2019-33-new}, and the specific implementation in this search is summarised below.
 
A loose, background-dominated selection is defined, denoted the \enquote{generation region} (GR), where the nominal isolation requirements from the SR are removed and the meson candidate $\pt$ requirement is relaxed to $\pt > 25$~\GeV.
The GR contains approximately 9700 events for the $D^0\gamma$ final states, and 1600 events for the $K^0_s\gamma$ final state.
From the events in the GR, pdfs are constructed to describe the distributions of the relevant kinematic and isolation variables and their most important correlations.
Pseudo-events, composed of a four momentum vector and  isolation variable for both the meson and photon, are generated from these pdfs using the ancestral sampling procedure described in Figure~\ref{fig:bkgdDiagram}.
In this diagram, the labels \enquote{1D} and \enquote{2D} refer to the dimensionality of the pdfs used to draw the associated variables.
Where two boxes in the diagram share a border, the values of the corresponding variables are sampled simultaneously from a combined pdf,
and arrows leading into boxes show that the variable is drawn from a pdf described in bins of the input variable generated in a previous stage in the sampling.
Vertices in the diagram labelled \enquote{Sum} show that the output variable, denoted by the arrow leading out of the vertex, is calculated as a sum of the input variables,
denoted by the dashed lines leading in to the vertex.
The nominal SR requirements are imposed on the ensemble of pseudo-events generated by this scheme, and the surviving pseudo-events are used to construct finely binned templates for the $m_{\cal{M}\gamma}$ distributions in each search, the discriminating variable used in the likelihood fit to the data.
These templates are smoothed using Gaussian kernel density estimation~\cite{Cranmer:2000du}.
 
\begin{figure}[!htbp]
\centering
\includegraphics[width=0.9\textwidth]{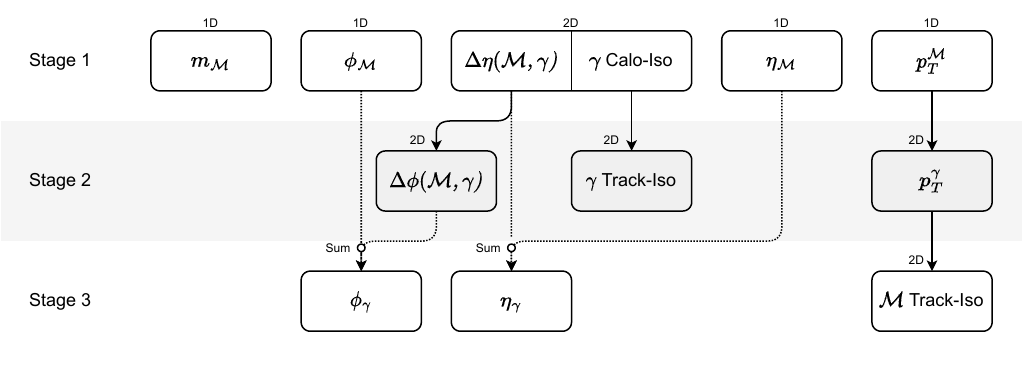}
\caption{The ancestral sampling scheme used in the non-parametric data-driven model for each final state, see text for details. Solid arrows indicate that the output variable is sampled in bins of the input variable, whilst dashed arrows indicate that the output variable is calculated as a sum of the input variables.\label{fig:bkgdDiagram}}
\end{figure}

The possibility of residual mismodelling of the background is addressed in the context of the maximum-likelihood fit described in Section~\ref{sec:result}.
The normalisation of the background model is determined directly from a fit to the observed data in the SR\@. To allow the model to adjust its shape
in the context of the fit to the $m_{\cal{M}\gamma}$ distributions of the data, three variations to the nominal model are derived which show notably distinct shapes in the $m_{\cal{M}\gamma}$ distribution. The first two alternative models are associated with distortions of the model's internal histograms for two kinematic variables,
namely the photon candidate $\pt$ and the angle $\Delta\phi$ between the pseudocandidate photon and meson. The form of the chosen distortions is motivated to introduce a
large enough shape variation to the resulting $m_{\cal{M}\gamma}$ template, ensuring the flexibility of the model to accommodate even large residual shape differences with the data.
The third alternative model is obtained by a direct transformation of the $m_{\cal{M}\gamma}$ template, based on a reweighting function with a linear dependence on $m_{\cal{M}\gamma}$.
This form is motivated by the general shape of the small residual pre-fit mismodelling observed in some earlier applications of the method~\cite{HIGG-2014-03,HIGG-2016-23,HDBS-2018-53-new,HIGG-2016-05,HIGG-2016-13_werratum,HDBS-2019-33-new}.
The technique of moment morphing is applied to parameterise the adjustment of the background prediction's shape between that of the nominal model and each variation~\cite{Baak:2014fta}.
Each of the three shape variations  has a dedicated nuisance parameter. To help stabilize the fit described in Section~\ref{sec:result},
the two nuisance parameters associated with the $\Delta\phi$ and photon $\pt$ distortions are each constrained by a separate Gaussian term in the likelihood.
However, the data in the SR has enough statistical power that all three nuisance parameters are well constrained in the fit to values associated with shape variations much milder than those of the three alternative models.

The just described model was validated through the definition of three intermediate event selection requirements, denoted validation regions (VR). Each validation region is defined by the loose GR selection with one key additional requirement taken from the tight SR selection, as outlined in Table~\ref{tab:regions}. The background modelling process was used to derive predictions for each of the three VRs, which were found to provide a good description of corresponding data samples, as shown in Figure~\ref{fig:bkgdVRdzero} for the $D^{0}\gamma$ final state and Figure~\ref{fig:bkgdVRkshort} for the $K_s^0\gamma$ final state.
\begin{table}[!htbp]
\centering
\caption{Definition of validation regions, each of which is based on the GR selection with a single additional requirement associated with the SR selection. In the case of the meson $\pt$ requirement, the value for the $K_s^0\gamma$ final state is shown in parentheses, while the value for the $D^0\gamma$ final state is not. The isolation requirements of all regions are common to both the $D^0\gamma$ and $K_s^0\gamma$ final states. }\label{tab:regions}
\begin{tabular}{lcccc}
\hline
Selection & Meson $\pt$ & Meson Isolation & Photon Isolation \\
\hline
GR & $>25\,\GeV$ & None & None \\
VR1 & $>39(38)\,\GeV$ & None & None \\
VR2 & $>25\,\GeV$ & Tight & None \\
VR3 & $>25\,\GeV$ & None & Tight \\
SR & $>39(38)\,\GeV$ & Tight & Tight \\
\hline
\end{tabular}
\end{table}
To further validate the model, the background modelling process was repeated with three additional independent validation regions, built using the same sampling scheme and including the shape variations.
These validation regions were identical to the nominal SR up to a single change in one of the requirements, resulting in event samples orthogonal to the SR\@.
One pair of validation regions, defined for both the $D^0\gamma$ and $K_s^0\gamma$ final states, involved a prompt-vertex selection region, in which the $L_{xy}$ significance requirement is inverted.
Another pair of validation regions considered events in which the two ID tracks that form the meson candidate are required to have the same sign charge, rather than be oppositely charged.
One further validation region was defined for the $K_s^0\gamma$ final state which selected $K^0_s$ candidates in a sideband of the mass peak, by requiring $538$~\MeV $< m_{K^0_s} < 1075$~\MeV.
In all cases, the background modelling procedure, when applied to these validation regions, exhibited good agreement with the data and served as a futher validation of the procedure.
Studies indicate that potential signal contamination in the GR would not be expected to result in a significant bias in the background prediction in the SR~\cite{Chisholm:2021pdn}.

\begin{figure}[!htbp]
\begin{center}
\subfigure[]{\includegraphics[width=0.32\textwidth]{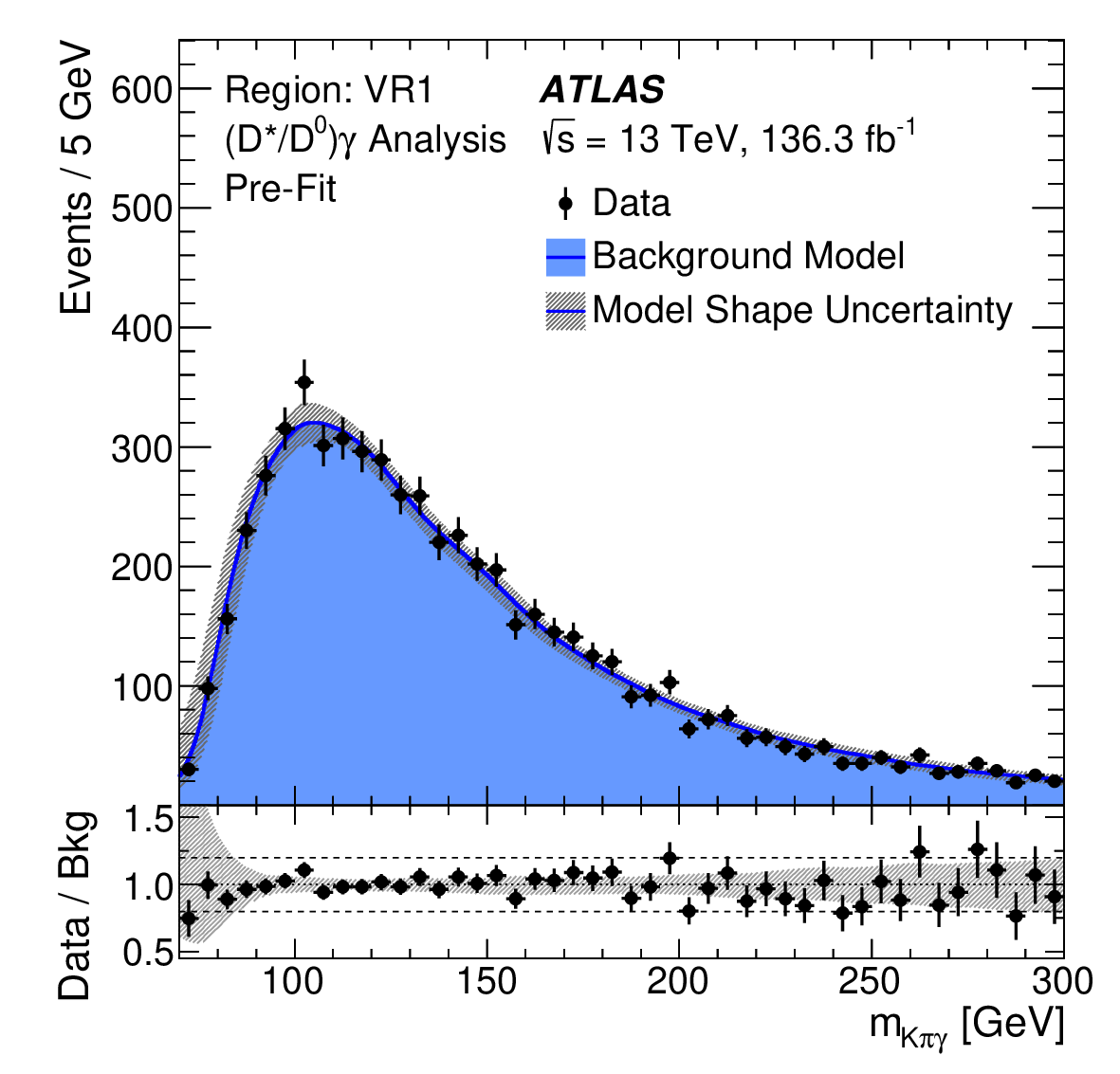}\label{fig:dzero:bkgdValVR1}}
\subfigure[]{\includegraphics[width=0.32\textwidth]{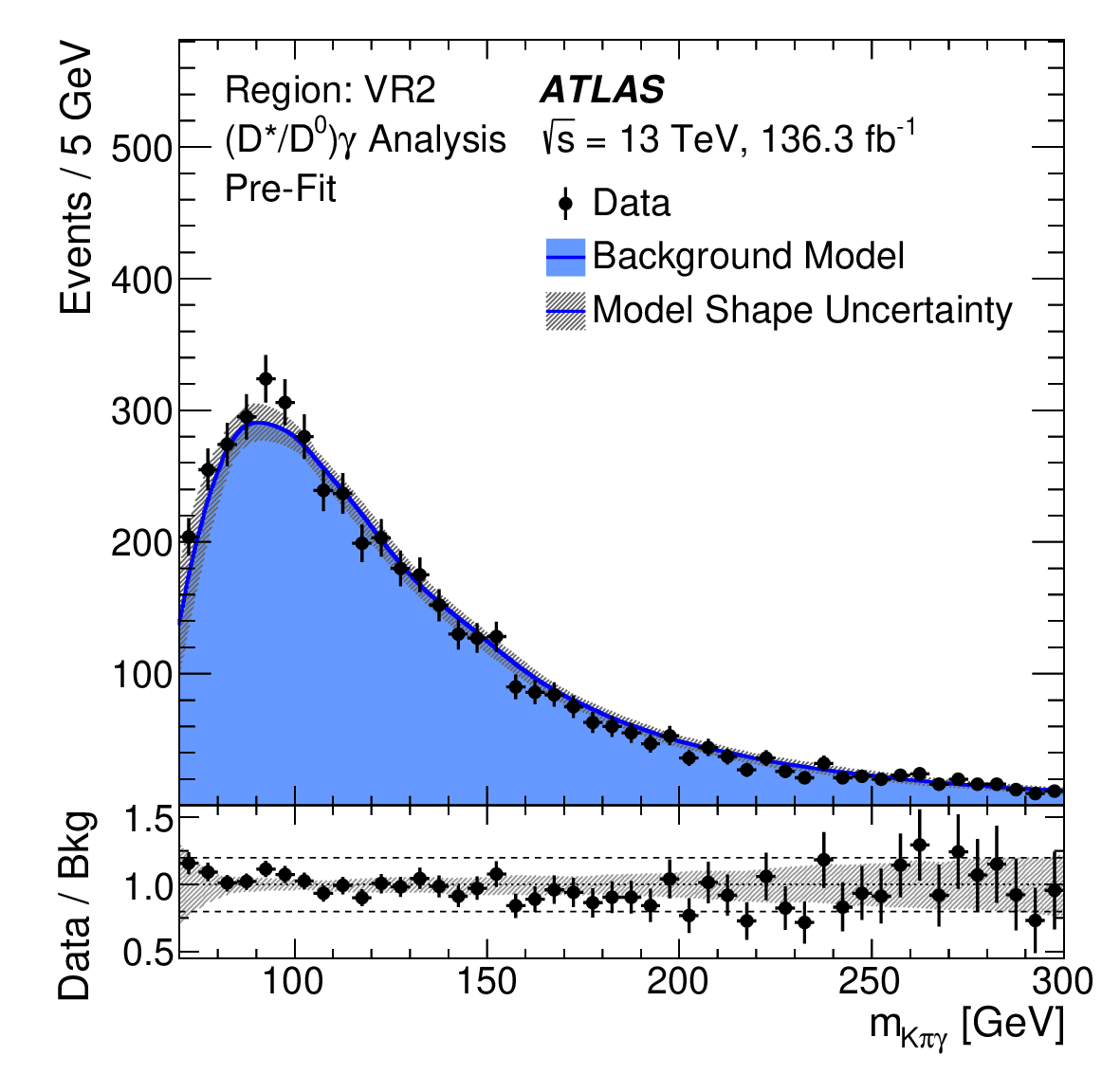}\label{fig:dzero:bkgdValVR2}}
\subfigure[]{\includegraphics[width=0.32\textwidth]{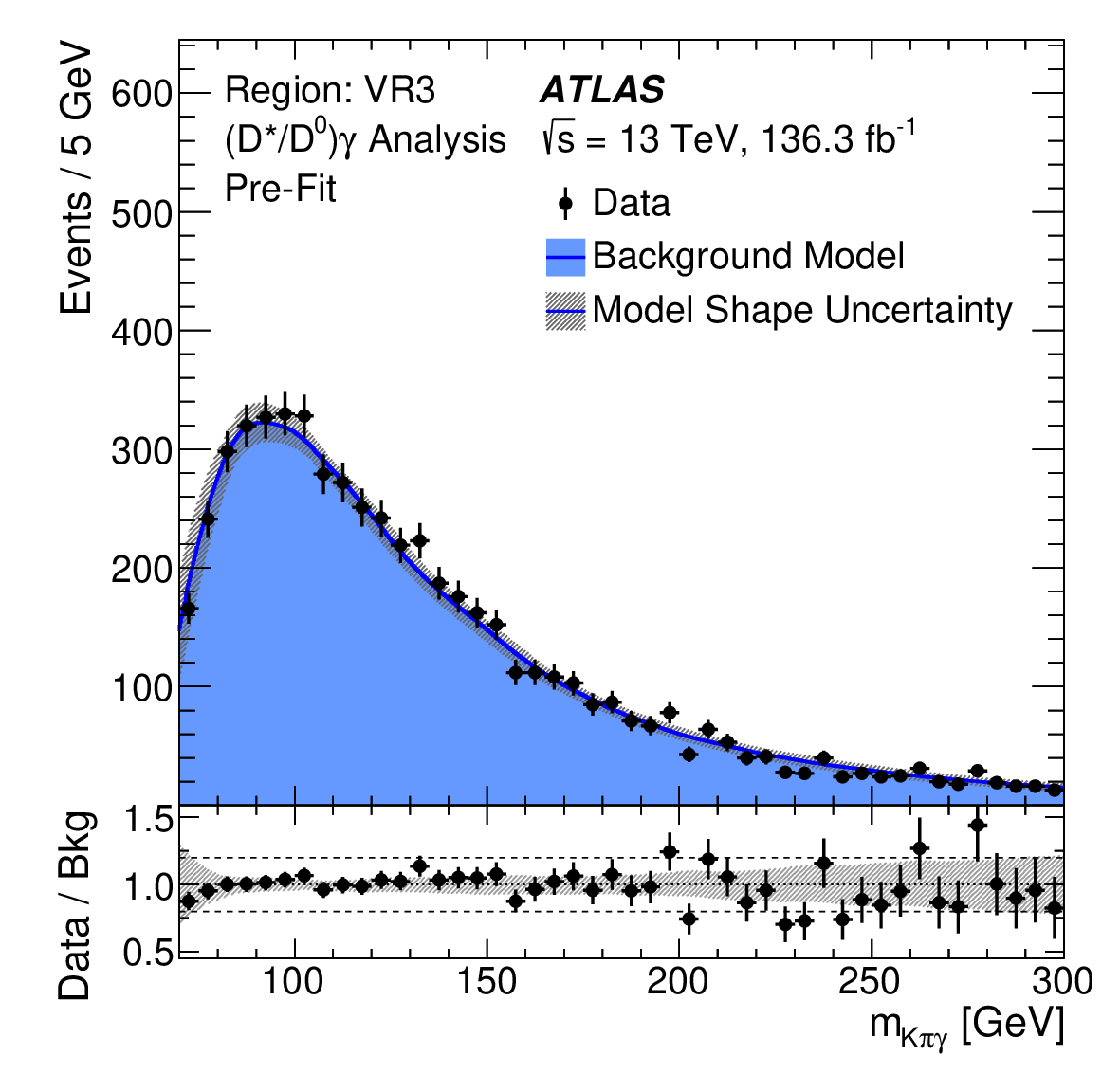}\label{fig:dzero:bkgdValVR3}}
 
\caption{The $m_{K^{\pm}\pi^{\mp}\gamma}$ distribution of the data and pre-fit background model prediction of validation regions for the $D^0\gamma$ final state. VR1 is shown in~\subref{fig:dzero:bkgdValVR1}, where the meson candidate is required to satisfy $\pt >39$~\GeV in addition to the GR requirements. VR2 is shown in~\subref{fig:dzero:bkgdValVR2}, where the meson-candidate isolation requirement used in the SR is applied, in addition to the GR requirements. VR3 is shown in~\subref{fig:dzero:bkgdValVR3}, where the photon-candidate isolation requirements used in the SR are applied, in addition to the GR requirements. In all cases, the ratio of the data to the prediction of the background model is shown in the lower panel.
\label{fig:bkgdVRdzero}}
\end{center}
\end{figure}
 
\begin{figure}[!htbp]
\begin{center}
\subfigure[]{\includegraphics[width=0.32\textwidth]{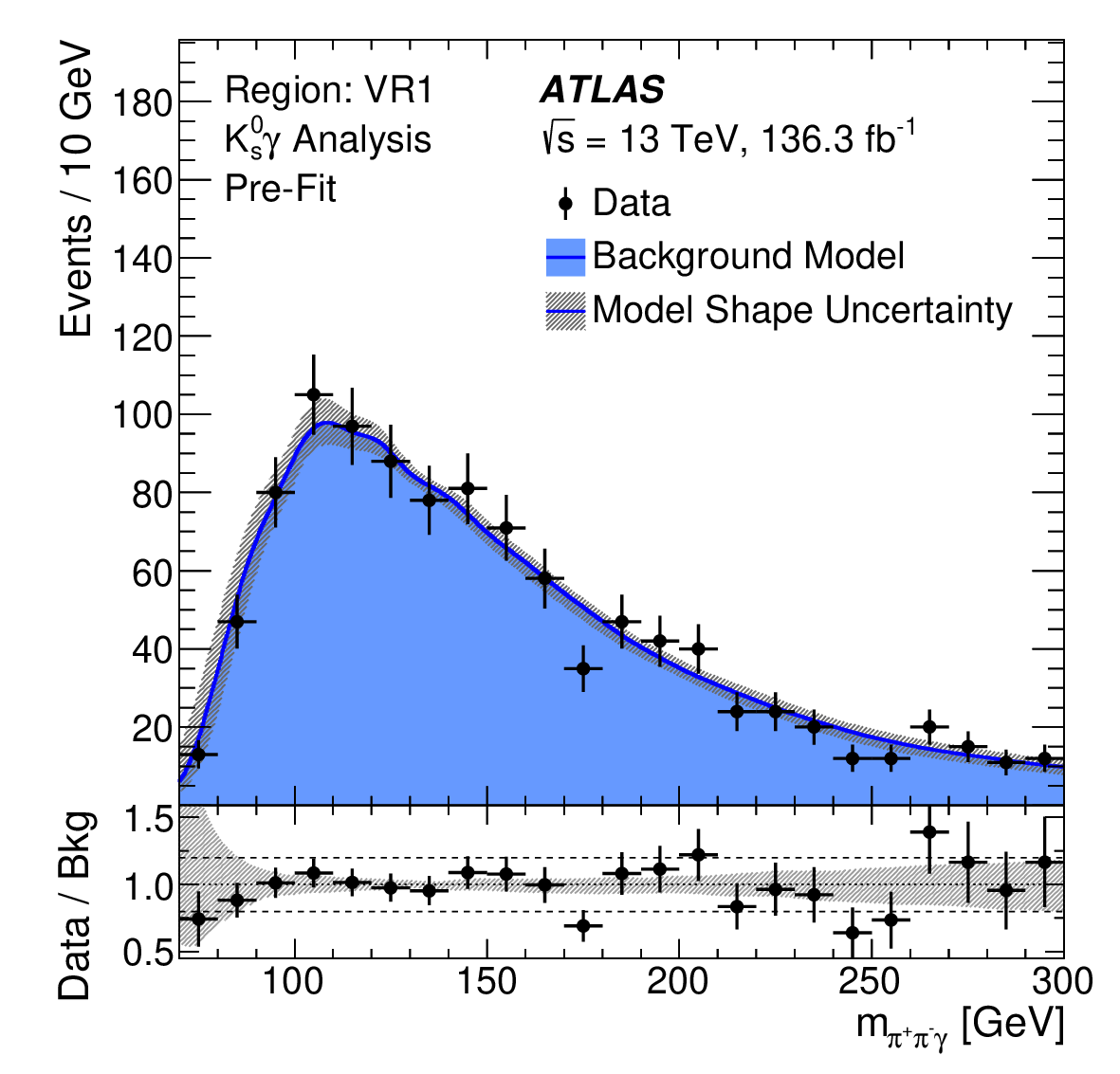}\label{fig:kshort:bkgdValVR1}}
\subfigure[]{\includegraphics[width=0.32\textwidth]{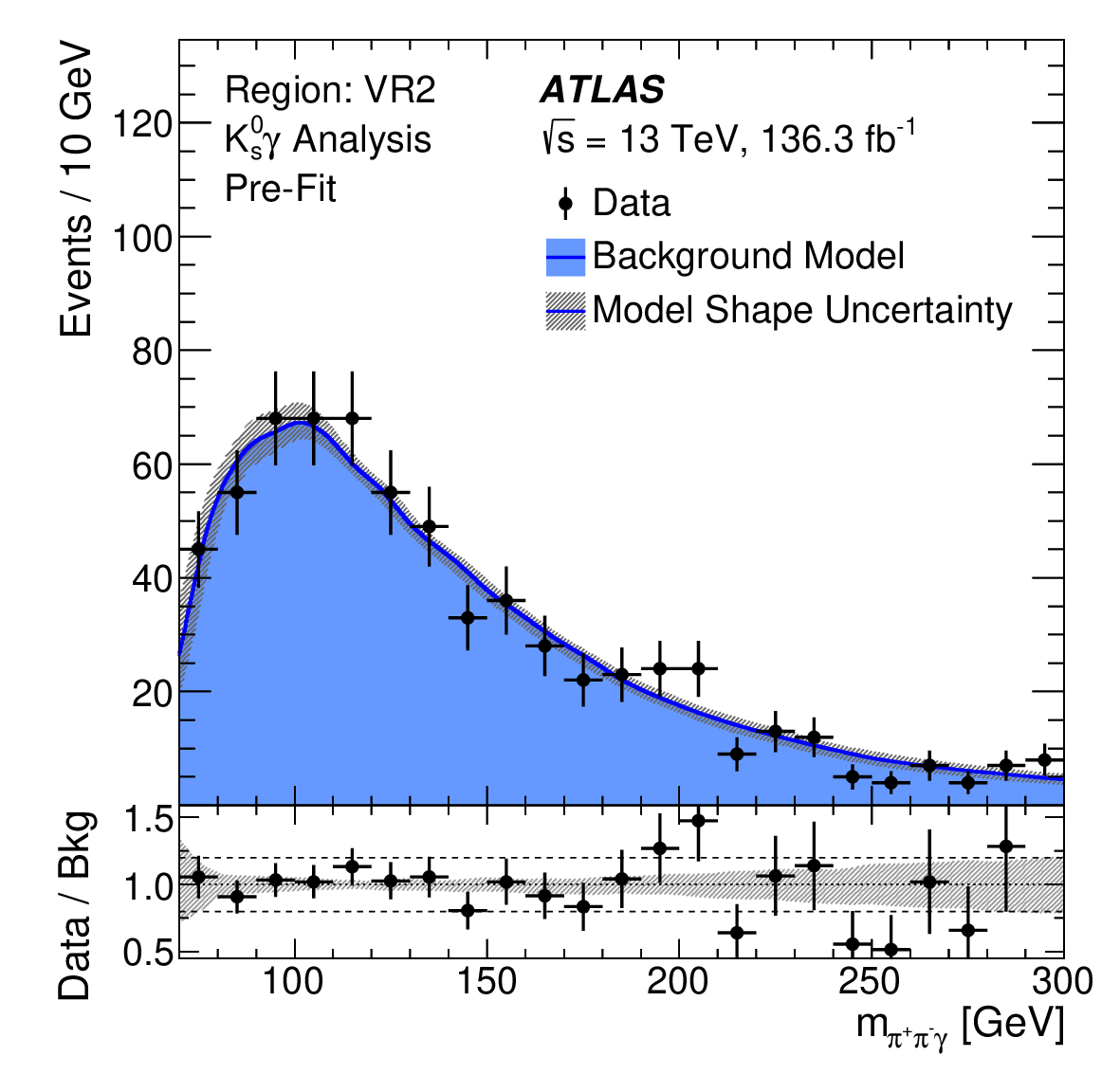}\label{fig:kshort:bkgdValVR2}}
\subfigure[]{\includegraphics[width=0.32\textwidth]{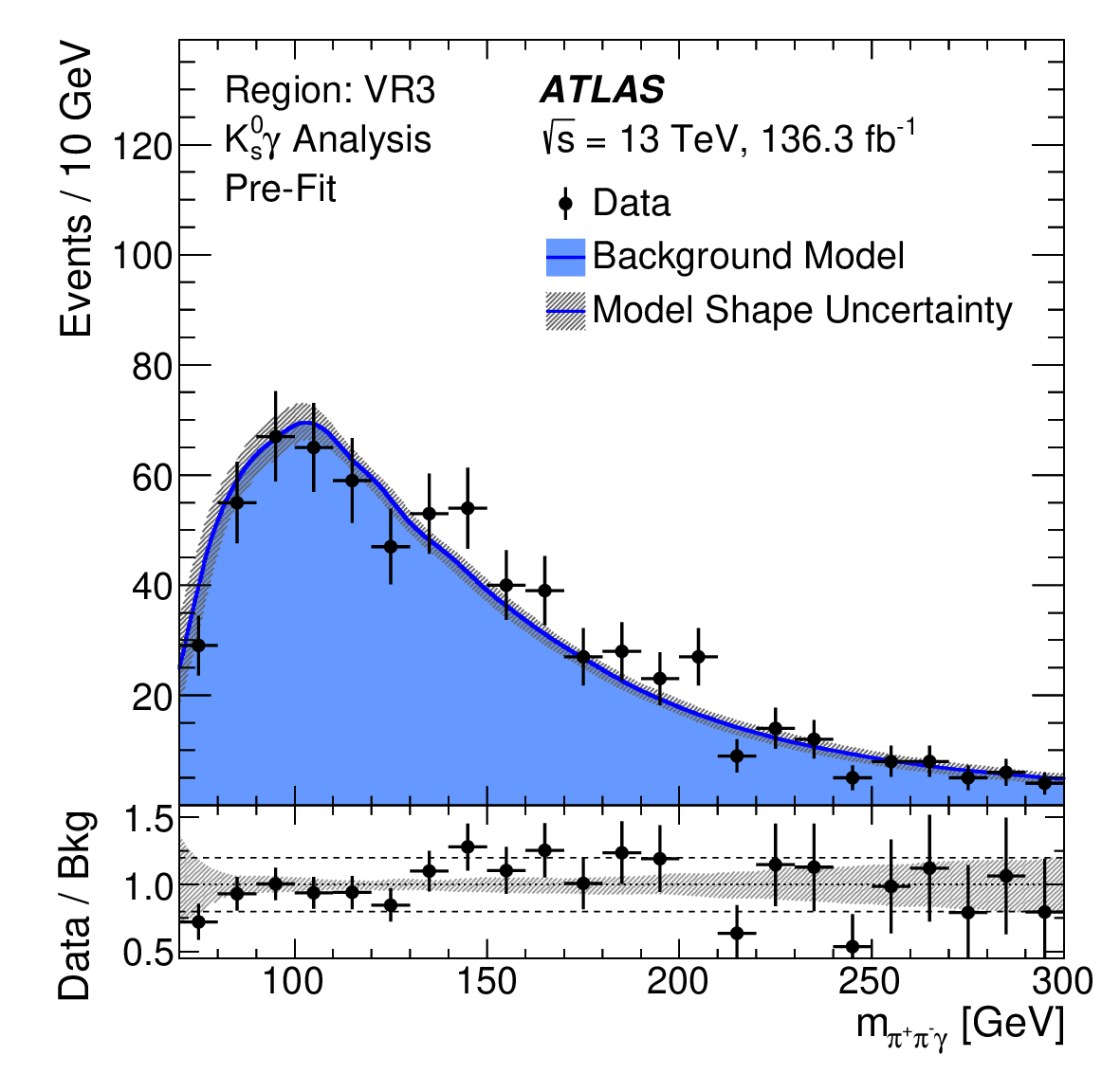}\label{fig:kshort:bkgdValVR3}}
 
\caption{The $m_{\pi^{+}\pi^{-}\gamma}$ distribution of the data and pre-fit background model prediction of validation regions for the $K^0_s\gamma$ final state. VR1 is shown in~\subref{fig:kshort:bkgdValVR1}, where the meson candidate is required to satisfy $\pt >38$~\GeV, in addition to  the GR requirements. VR2 is shown in~\subref{fig:kshort:bkgdValVR2}, where the meson-candidate isolation requirement used in the SR is applied, in addition to the GR requirements. VR3 is shown in~\subref{fig:kshort:bkgdValVR3}, where the photon-candidate isolation requirements used in the SR are applied, in addition to the GR requirements. In all cases, the ratio of the data to the prediction of the background model is shown in the lower panel.
\label{fig:bkgdVRkshort}}
\end{center}
\end{figure}


\section{Systematic uncertainties}
\label{sec:systematics}
 
Systematic uncertainties in the signal yield and inferred branching fraction of the Higgs and $Z$ boson decays are considered.
Uncertainties in the Higgs boson production cross sections from the QCD scale total $5.0\%$ and from the PDFs and strong coupling constant, $\alphas$, total $3.2\%$~\cite{deFlorian:2016spz,Dainese:2019rgk}.
 
The corresponding uncertainty in the $Z$ boson production cross-section is $2.9\%$~\cite{STDM-2015-03}.
The estimated integrated luminosity has an uncertainty of 0.84\%, which is calculated using the method described in Ref.~\cite{DAPR-2021-01}, and uses the LUCID-2 detector~\cite{LUCID2} for the primary luminosity measurements, complemented by measurements using the inner detector and calorimeters.
 
The uncertainty in the acceptance of the Higgs boson signal is estimated by varying the QCD normalisation and factorisation scales, $\alphas$, PDFs, set of tuned parameters for the underlying event, and parton showering at generator level.
The total uncertainty in the $H\to D^*\gamma$ signal acceptance is estimated to be 2.0\%.
For the $Z$ boson, the respective signal acceptance uncertainty is determined to be 0.40\%--0.52\% by comparing the $Z$ boson kinematic distributions in simulated events with measurements in data~\cite{STDM-2018-14}.
 
Trigger efficiencies for photons are determined from samples enriched with $Z\to\ee$ events in data~\cite{TRIG-2018-05}.
The photon trigger efficiency is estimated to contribute a systematic uncertainty of 0.66\%--0.72\% to each of the expected signal yields~\cite{TRIG-2018-05,TRIG-2018-01}.
The uncertainty in the track component of the trigger is derived by varying the leading track $\et/\pt$ distribution based on comparisons between data measurements and simulated events in Ref.~\cite{JETM-2018-08}, and is $2.9\%$ for $H\to D^*\gamma$, $3.9\%$ for $Z\to D^0\gamma$, and $6.0\%$ for $Z\to K^0_s\gamma$.
Photon identification efficiencies are determined using the enriched $Z\to\ee$ event samples, and inclusive photon events and $Z\to\ell^+\ell^-\gamma$ events~\cite{EGAM-2018-01}.
The photon identification efficiency uncertainties are 1.6\%--1.9\% for the Higgs and $Z$~boson signals.
The photon energy scale uncertainty, determined from $Z\to\ee$ events and validated using $Z\to\ell^+\ell^- \gamma$ events~\cite{EGAM-2018-01}, is propagated through the simulated samples as a function of $\etagamma$ and $\ptgamma$.
 
The track reconstruction efficiency uncertainties total $1.0\%$. 
The uncertainty in the photon energy scale and in the resolution in the simulation has a 0.12\%--0.44\% effect on the Higgs and $Z$ boson signal yields.
To assess any effect on the expected signal yield from imperfect modelling of pile-up, the average number of pile-up interactions is varied in the simulation; the corresponding uncertainty is 2.9\%--3.1\%.
The uncertainty in the efficiency to reconstruct decays with displaced vertices is estimated by using the studies in Ref.~\cite{PERF-2015-07},
and is $1.5\%$ for the $D^0$ mesons in $H\to D^*\gamma$ and $Z\to D^0\gamma$, and $3.5\%$ for the $K^0_s$ mesons in $Z\to K^0_s\gamma$.
 
Combining the individual sources of uncertainty gives a  total uncertainty in the signal yields equal to $7.9\%$ for the $H\to D^*\gamma$ signal, $6.4\%$ for $Z\to D^0\gamma$, and $8.5\%$ for $Z\to K^0_s\gamma$.
The uncertainty in the shape from each of the considered signal systematic uncertainties and from the statistical uncertainty in the simulated samples was found to be negligible.
Systematic uncertainties in the shape of the background model are also considered, as described in Section~\ref{sec:background}.


 
\section{Results}
\label{sec:result}
 
In total, 2243 events are observed in the signal region for the $D^0\gamma$ and $D^*\gamma$ final states and 283 events are observed in the signal region for the $K^0_s\gamma$ final state.
The selected data are compared with background and signal predictions using an unbinned maximum-likelihood fit to the $m_{\mathcal{M}\gamma}$ distribution in the range of $70$~\GeV $< m_{\mathcal{M}\gamma} < 300$~\GeV.
The parameters of interest are the Higgs and $Z$ boson signal normalisations, and systematic uncertainties are modelled using additional nuisance parameters in the fit.
The background normalisation is a free parameter in the model.
Upper limits are set on the branching fractions for each of the Higgs and $Z$ boson signal decays using the CL$_\text{s}$ modified frequentist formalism~\cite{Read:2002hq} with the profile-likelihood-ratio test statistic and the asymptotic approximations derived in Ref.~\cite{Cowan:2010js}. For $D^0\gamma$ and $D^*\gamma$ final states, limits on the $H\rightarrow D^*\gamma$ signal are obtained while profiling the signal normalisation parameter of the $Z\rightarrow D^0\gamma$ process (and vice versa).
The SM production cross-section is assumed for the Higgs boson while the ATLAS measurement of the inclusive $Z$ boson cross-section is used for the $Z$ boson signals, as discussed in Section~\ref{sec:signal}.
 
The results of the background-only fits for each search are shown in Figure~\ref{fig:fits},
where the signal distributions shown correspond to the extracted 95\% confidence-level (CL) branching fraction upper limits.
The expected and observed numbers of background events in the three-body mass ranges relevant to the Higgs and $Z$ boson signals are given in Table~\ref{tab:events},
where the expected backgrounds are obtained from fits to these backgrounds.
Table~\ref{tab:events} also shows the expected number of signal events for reference branching fractions near the sensitivity of the analysis: $10^{-3}$ for $H\rightarrow D^*\gamma$ and $10^{-6}$ for $Z\rightarrow D^0\gamma$ and $Z\rightarrow K^0_s\gamma$. No significant signal is observed in any of the search channels and the $p$-values for the background only hypothesis are $0.24$, $0.23$ and $0.43$ in the $H\rightarrow D^*\gamma$, $Z\rightarrow D^0\gamma$ and $Z\rightarrow K^0_s\gamma$ channels, respectively.
 
\begin{figure}[!htbp]
\centering
\subfigure[]{\includegraphics[width=0.49\textwidth]{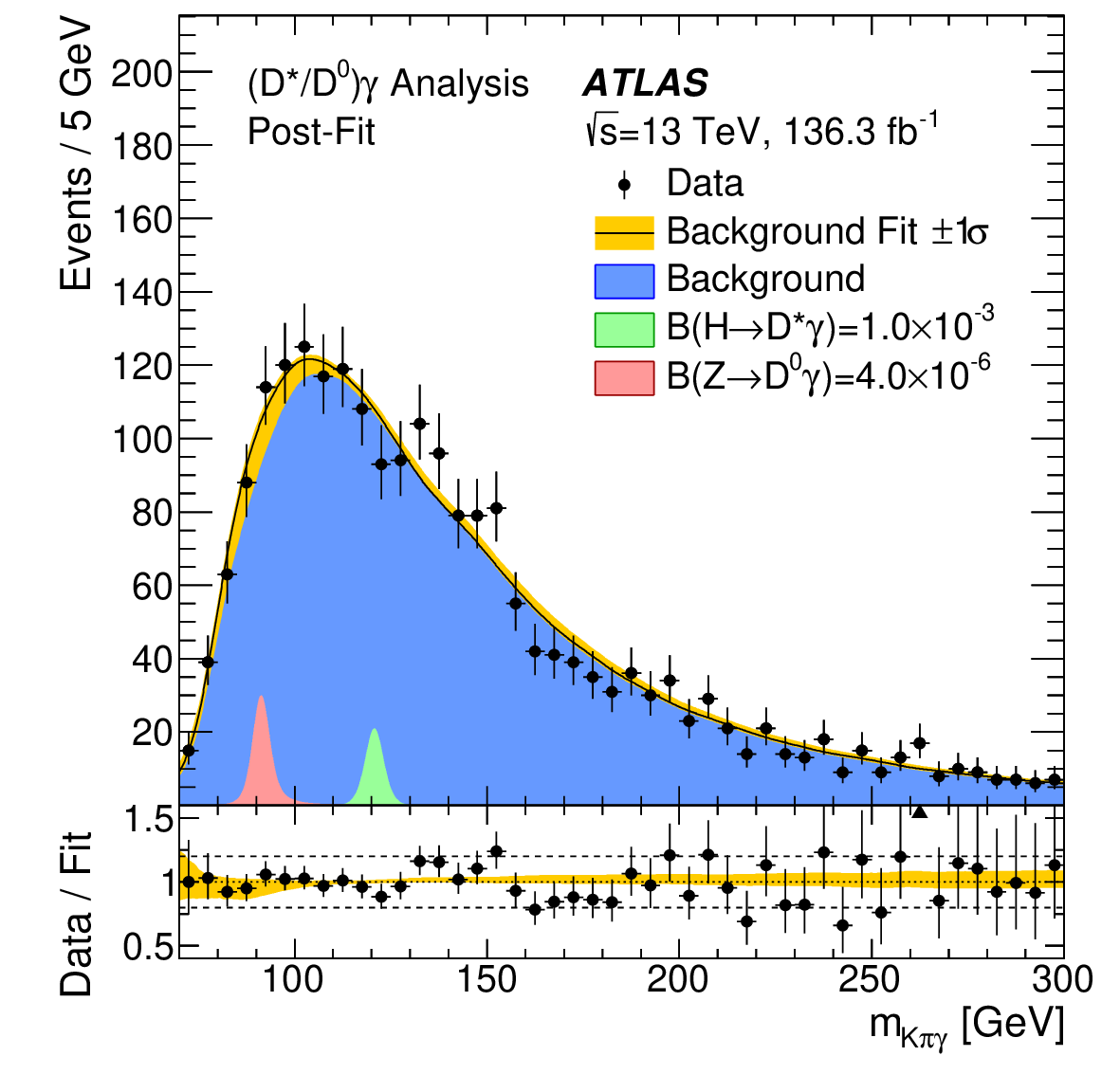}\label{fig:DZero}}
\subfigure[]{\includegraphics[width=0.49\textwidth]{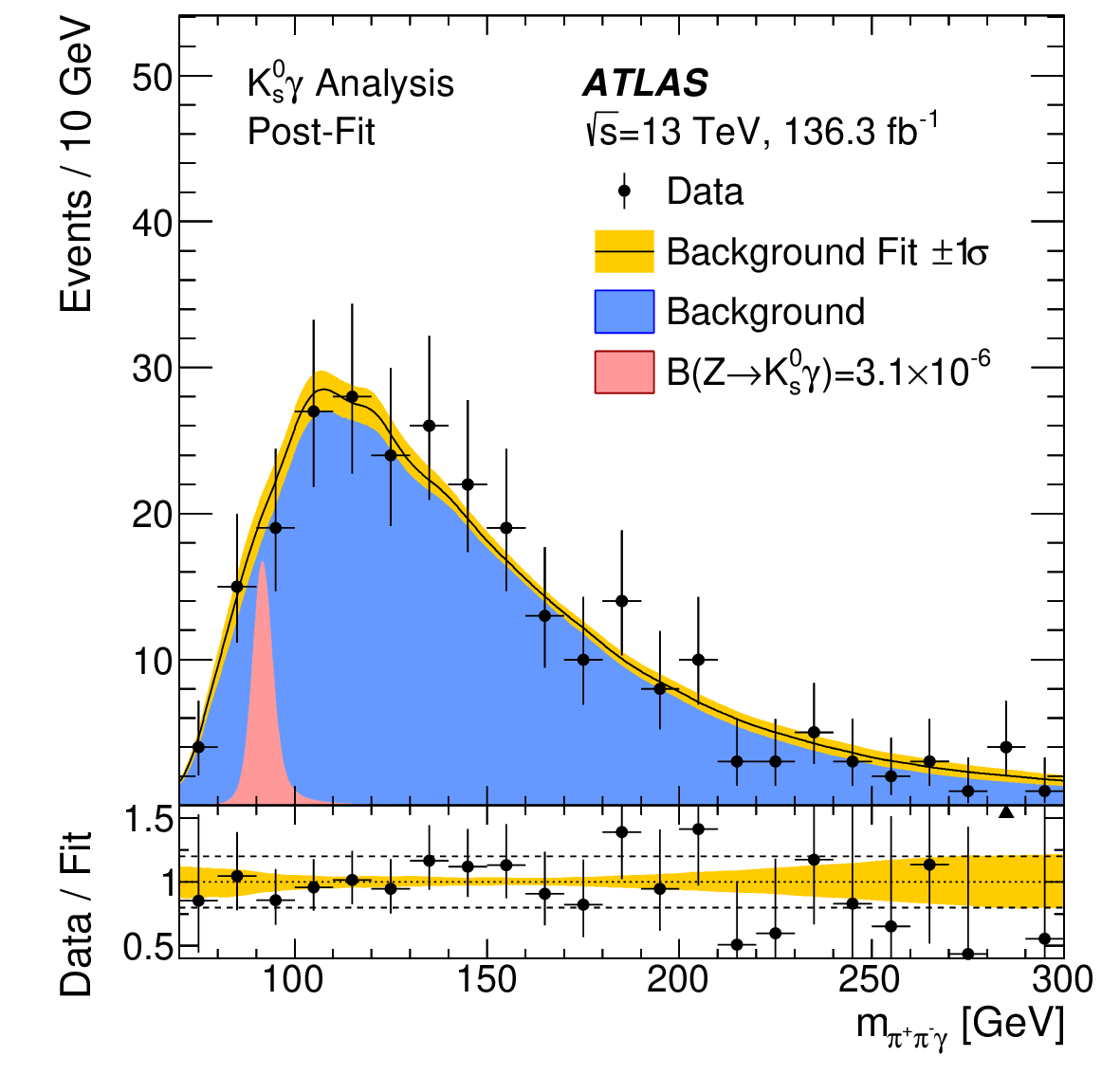}\label{fig:KShort}}
\caption{Comparison between data and background prediction for the $m_{\mathcal{M}\gamma}$ distribution after the background-only fit (``Post-Fit'') in the signal region for the \subref{fig:DZero} $D^*\gamma$ and $D^0\gamma$ final states and for the \subref{fig:KShort} $K^0_s\gamma$ final state.
The unbinned background pdf is shown with a yellow band that represents the uncertainty in the fit arising from the constrained background shape systematic uncertainties. This uncertainty is largest in the region $m_{\mathcal{M}\gamma}< 100\,\GeV$, where the gradient of the distributions vary most. The choice of bin width is motivated only to best visualise the data and is independent of the likelihood definition. The lower panels show the ratio between the data and the background prediction. The expected signal distributions are shown normalised to a branching fraction corresponding to the observed $95\%$ CL upper limit. }
\label{fig:fits}
\end{figure}
 
\begin{table}[!htbp]
\centering
\caption{Numbers of observed and expected background events for the $m_{\mathcal{M}\gamma}$ ranges of interest.
Each expected background and the corresponding uncertainty is obtained by integrating the total pdf after a background-only fit to the data, where the uncertainty does not take into account statistical fluctuations in each mass range.
Expected Higgs and $Z$ boson signal contributions, with their corresponding total systematic uncertainty, are shown for reference branching fractions of $10^{-3}$ and $10^{-6}$, respectively.
Entries are marked with a dash when there is no signal of that type in the specified range.}\label{tab:events}
\begin{tabular}{lcccc}
\hline
Channel            & Mass range  & Observed (Expected)  & $H$ signal & $Z$ signal  \\
& $[\GeV]$ & background &  ${\cal B}=10^{-3}$ & ${\cal B}=10^{-6}$ \\ \hline
$H\to D^*\gamma$ & 116--126 &            203 $(214.8 \pm 5.5)$  & 25.4 $\pm$ 2.0 & --          \\
$Z\to D^0\gamma$ & 86--96   & \phantom{0}215 $(206 \pm 14)$ & --             & 10.3 $\pm$ 0.7 \\
$Z\to K^0_s\gamma$ & 86--96 &             21 $(19.5 \pm 2.0)$   & --             & \phantom{0}4.2 $\pm$ 0.4 \\  \hline
\end{tabular}
\end{table}
 
The expected and observed 95\% CL upper limits on the branching fractions for Higgs and $Z$ boson decays into each final state are presented in Table~\ref{tab:limits}, along with the observed upper limits in terms of Higgs and $Z$ boson production cross-section times branching fraction for each decay.
The systematic uncertainties in the signal normalisation and background shape described respectively in
Sections~\ref{sec:signal} and~\ref{sec:background} result in a 2\% increase of the expected 95\% CL upper limit on the
branching fraction of the $H\to D^*\gamma$ decay, a 9\% increase for the $Z\to D^0\gamma$ decay, and a 10\% increase for the $Z\to K^0_s\gamma$ decay, mostly due to the systematic uncertainty in the background shape.
 
\begin{table}[!htbp]
\centering
\caption{Observed and expected (with the corresponding $\pm1\sigma$ intervals) 95$\%$ CL upper limits on the branching fractions for $H\rightarrow D^*\gamma$, $Z\rightarrow D^0\gamma$ and $Z\rightarrow K^0_s\gamma$.
Standard Model production of the Higgs boson is assumed.
The corresponding upper limits on the production cross-section times branching fraction $\sigma\times\mathcal{B}$ are also shown.}
\label{tab:limits}
\begin{tabular}{l|cc|cc}
\cline{2-5}
\multicolumn{1}{c}{} & \multicolumn{4}{c}{95\% CL upper limits} \\ \cline{2-5}
\multicolumn{1}{c}{} & \multicolumn{2}{c|}{Branching Fraction} & \multicolumn{2}{c}{$\sigma\times\mathcal{B}$~[fb]} \\\hline
Channel            & Observed               & Expected    &  Observed & Expected \\ \hline
$H\to D^*\gamma$  & $1.0\times 10^{-3}$ & $\tabscript{1.2}{+0.5}{-0.3}\times 10^{-3}$  & $\phantom{0}58$ & $\phantom{0}\tabscript{68}{+28}{-19}$  \\
$Z\to D^0\gamma$ & $4.0\times 10^{-6}$   & $\tabscript{3.4}{+1.4}{-1.0}\times 10^{-6}$ & $235$  & $\tabscript{200}{+82}{-56}$    \\
$Z\to K^0_s\gamma$ & $3.1\times 10^{-6}$ & $\tabscript{3.0}{+1.3}{-0.8}\times 10^{-6}$ & $185$  & $\tabscript{176}{+77}{-49}$  \\\hline
\end{tabular}
\end{table}


\FloatBarrier

\section{Conclusions}
\label{sec:summary}
 
Searches for the flavour-violating exclusive decays of the Higgs boson $H\rightarrow D^*\gamma$, $Z\rightarrow D^0\gamma$ and $Z\rightarrow K^0_s\gamma$ were made using $136.3~\ifb$ of $pp$ collision data at $\sqrt{s} = 13$~\TeV collected with the ATLAS detector at the LHC.
Each of these decay channels involve a displaced vertex, either from the decay of a $D^0$ meson or from a $K^0_s$ meson.
The observed data are compatible with the expected backgrounds. The observed $95\%$ CL upper limits are $\mathcal{B}(H\rightarrow D^*\gamma) < 1.0\times 10^{-3}$, $\mathcal{B}(Z\rightarrow D^0\gamma) < 4.0\times 10^{-6}$, and $\mathcal{B}(Z\rightarrow K^0_s\gamma) < 3.1\times 10^{-6}$. The corresponding expected $95\%$ CL upper limits on the branching fractions of each decay are $\mathcal{B}(H\rightarrow D^*\gamma) < 1.2\times 10^{-3}$, $\mathcal{B}(Z\rightarrow D^0\gamma) < 3.4\times 10^{-6}$, and $\mathcal{B}(Z\rightarrow K^0_s\gamma) < 3.0\times 10^{-6}$, where Standard Model Higgs boson production is assumed. These results represent the first limits set on the decays $H\rightarrow D^*\gamma$ and $Z\rightarrow K^0_s\gamma$, and a factor of approximately 500 improvement on the $Z\rightarrow D^0\gamma$ limit set by LHCb.


\section*{Acknowledgements}


We thank CERN for the very successful operation of the LHC and its injectors, as well as the support staff at
CERN and at our institutions worldwide without whom ATLAS could not be operated efficiently.
 
The crucial computing support from all WLCG partners is acknowledged gratefully, in particular from CERN, the ATLAS Tier-1 facilities at TRIUMF/SFU (Canada), NDGF (Denmark, Norway, Sweden), CC-IN2P3 (France), KIT/GridKA (Germany), INFN-CNAF (Italy), NL-T1 (Netherlands), PIC (Spain), RAL (UK) and BNL (USA), the Tier-2 facilities worldwide and large non-WLCG resource providers. Major contributors of computing resources are listed in Ref.~\cite{ATL-SOFT-PUB-2023-001}.
 
We gratefully acknowledge the support of ANPCyT, Argentina; YerPhI, Armenia; ARC, Australia; BMWFW and FWF, Austria; ANAS, Azerbaijan; CNPq and FAPESP, Brazil; NSERC, NRC and CFI, Canada; CERN; ANID, Chile; CAS, MOST and NSFC, China; Minciencias, Colombia; MEYS CR, Czech Republic; DNRF and DNSRC, Denmark; IN2P3-CNRS and CEA-DRF/IRFU, France; SRNSFG, Georgia; BMBF, HGF and MPG, Germany; GSRI, Greece; RGC and Hong Kong SAR, China; ISF and Benoziyo Center, Israel; INFN, Italy; MEXT and JSPS, Japan; CNRST, Morocco; NWO, Netherlands; RCN, Norway; MNiSW, Poland; FCT, Portugal; MNE/IFA, Romania; MESTD, Serbia; MSSR, Slovakia; ARRS and MIZ\v{S}, Slovenia; DSI/NRF, South Africa; MICINN, Spain; SRC and Wallenberg Foundation, Sweden; SERI, SNSF and Cantons of Bern and Geneva, Switzerland; MOST, Taipei; TENMAK, T\"urkiye; STFC, United Kingdom; DOE and NSF, United States of America.
 
Individual groups and members have received support from BCKDF, CANARIE, CRC and DRAC, Canada; CERN-CZ, PRIMUS 21/SCI/017 and UNCE SCI/013, Czech Republic; COST, ERC, ERDF, Horizon 2020, ICSC-NextGenerationEU and Marie Sk{\l}odowska-Curie Actions, European Union; Investissements d'Avenir Labex, Investissements d'Avenir Idex and ANR, France; DFG and AvH Foundation, Germany; Herakleitos, Thales and Aristeia programmes co-financed by EU-ESF and the Greek NSRF, Greece; BSF-NSF and MINERVA, Israel; Norwegian Financial Mechanism 2014-2021, Norway; NCN and NAWA, Poland; La Caixa Banking Foundation, CERCA Programme Generalitat de Catalunya and PROMETEO and GenT Programmes Generalitat Valenciana, Spain; G\"{o}ran Gustafssons Stiftelse, Sweden; The Royal Society and Leverhulme Trust, United Kingdom.
 
In addition, individual members wish to acknowledge support from CERN: European Organization for Nuclear Research (CERN PJAS); Chile: Agencia Nacional de Investigaci\'on y Desarrollo (FONDECYT 1190886, FONDECYT 1210400, FONDECYT 1230812, FONDECYT 1230987); China: National Natural Science Foundation of China (NSFC - 12175119, NSFC 12275265, NSFC-12075060); Czech Republic: PRIMUS Research Programme (PRIMUS/21/SCI/017); European Union: European Research Council (ERC - 948254, ERC 101089007), Horizon 2020 Framework Programme (MUCCA - CHIST-ERA-19-XAI-00), European Union, Future Artificial Intelligence Research (FAIR-NextGenerationEU PE00000013), Italian Center for High Performance Computing, Big Data and Quantum Computing (ICSC, NextGenerationEU); France: Agence Nationale de la Recherche (ANR-20-CE31-0013, ANR-21-CE31-0013, ANR-21-CE31-0022), Investissements d'Avenir Labex (ANR-11-LABX-0012); Germany: Baden-Württemberg Stiftung (BW Stiftung-Postdoc Eliteprogramme), Deutsche Forschungsgemeinschaft (DFG - 469666862, DFG - CR 312/5-2); Italy: Istituto Nazionale di Fisica Nucleare (ICSC, NextGenerationEU); Japan: Japan Society for the Promotion of Science (JSPS KAKENHI JP21H05085, JSPS KAKENHI JP22H01227, JSPS KAKENHI JP22H04944, JSPS KAKENHI JP22KK0227); Netherlands: Netherlands Organisation for Scientific Research (NWO Veni 2020 - VI.Veni.202.179); Norway: Research Council of Norway (RCN-314472); Poland: Polish National Agency for Academic Exchange (PPN/PPO/2020/1/00002/U/00001), Polish National Science Centre (NCN 2021/42/E/ST2/00350, NCN OPUS nr 2022/47/B/ST2/03059, NCN UMO-2019/34/E/ST2/00393, UMO-2020/37/B/ST2/01043, UMO-2021/40/C/ST2/00187, UMO-2022/47/O/ST2/00148); Slovenia: Slovenian Research Agency (ARIS grant J1-3010); Spain: BBVA Foundation (LEO22-1-603), Generalitat Valenciana (Artemisa, FEDER, IDIFEDER/2018/048), Ministry of Science and Innovation (MCIN \& NextGenEU PCI2022-135018-2, MICIN \& FEDER PID2021-125273NB, RYC2019-028510-I, RYC2020-030254-I, RYC2021-031273-I, RYC2022-038164-I), PROMETEO and GenT Programmes Generalitat Valenciana (CIDEGENT/2019/023, CIDEGENT/2019/027); Sweden: Swedish Research Council (VR 2018-00482, VR 2022-03845, VR 2022-04683, VR grant 2021-03651), Knut and Alice Wallenberg Foundation (KAW 2017.0100, KAW 2018.0157, KAW 2018.0458, KAW 2019.0447, KAW 2022.0358); Switzerland: Swiss National Science Foundation (SNSF - PCEFP2\_194658); United Kingdom: Leverhulme Trust (Leverhulme Trust RPG-2020-004), Royal Society (NIF-R1-231091); United States of America: U.S. Department of Energy (ECA DE-AC02-76SF00515), Neubauer Family Foundation.


\printbibliography
 
\clearpage
 
\begin{flushleft}
\hypersetup{urlcolor=black}
{\Large The ATLAS Collaboration}

\bigskip

\AtlasOrcid[0000-0002-6665-4934]{G.~Aad}$^\textrm{\scriptsize 103}$,
\AtlasOrcid[0000-0001-7616-1554]{E.~Aakvaag}$^\textrm{\scriptsize 16}$,
\AtlasOrcid[0000-0002-5888-2734]{B.~Abbott}$^\textrm{\scriptsize 121}$,
\AtlasOrcid[0000-0002-1002-1652]{K.~Abeling}$^\textrm{\scriptsize 55}$,
\AtlasOrcid[0000-0001-5763-2760]{N.J.~Abicht}$^\textrm{\scriptsize 49}$,
\AtlasOrcid[0000-0002-8496-9294]{S.H.~Abidi}$^\textrm{\scriptsize 29}$,
\AtlasOrcid[0009-0003-6578-220X]{M.~Aboelela}$^\textrm{\scriptsize 44}$,
\AtlasOrcid[0000-0002-9987-2292]{A.~Aboulhorma}$^\textrm{\scriptsize 35e}$,
\AtlasOrcid[0000-0001-5329-6640]{H.~Abramowicz}$^\textrm{\scriptsize 152}$,
\AtlasOrcid[0000-0002-1599-2896]{H.~Abreu}$^\textrm{\scriptsize 151}$
% ...
\bigskip
\\

% ...
$^{16}$Department for Physics and Technology, University of Bergen, Bergen; Norway.\\
% ...
$^{29}$Physics Department, Brookhaven National Laboratory, Upton NY; United States of America.\\
% ...
$^{35}$$^{(a)}$Facult\'e des Sciences Ain Chock, R\'eseau Universitaire de Physique des Hautes Energies - Universit\'e Hassan II, Casablanca;$^{(b)}$Facult\'{e} des Sciences, Universit\'{e} Ibn-Tofail, K\'{e}nitra;$^{(c)}$Facult\'e des Sciences Semlalia, Universit\'e Cadi Ayyad, LPHEA-Marrakech;$^{(d)}$LPMR, Facult\'e des Sciences, Universit\'e Mohamed Premier, Oujda;$^{(e)}$Facult\'e des sciences, Universit\'e Mohammed V, Rabat;$^{(f)}$Institute of Applied Physics, Mohammed VI Polytechnic University, Ben Guerir; Morocco.\\
% ...
$^{44}$Physics Department, Southern Methodist University, Dallas TX; United States of America.\\
% ...
$^{49}$Fakult\"{a}t Physik , Technische Universit{\"a}t Dortmund, Dortmund; Germany.\\
% ...
$^{55}$II. Physikalisches Institut, Georg-August-Universit\"{a}t G\"ottingen, G\"ottingen; Germany.\\
% ...
$^{103}$CPPM, Aix-Marseille Universit\'e, CNRS/IN2P3, Marseille; France.\\
% ...
$^{121}$Homer L. Dodge Department of Physics and Astronomy, University of Oklahoma, Norman OK; United States of America.\\
% ...
$^{151}$Department of Physics, Technion, Israel Institute of Technology, Haifa; Israel.\\
$^{152}$Raymond and Beverly Sackler School of Physics and Astronomy, Tel Aviv University, Tel Aviv; Israel.\\
% ...
$^{a}$ Also Affiliated with an institute covered by a cooperation agreement with CERN.\\
$^{b}$ Also at An-Najah National University, Nablus; Palestine.\\
$^{c}$ Also at Borough of Manhattan Community College, City University of New York, New York NY; United States of America.\\
$^{d}$ Also at Center for High Energy Physics, Peking University; China.\\
$^{e}$ Also at Center for Interdisciplinary Research and Innovation (CIRI-AUTH), Thessaloniki; Greece.\\
$^{f}$ Also at Centro Studi e Ricerche Enrico Fermi; Italy.\\
% ...

\end{flushleft}


\end{document}